\documentclass[12pt]{article}
\usepackage[dvips]{graphicx}

\newcommand{\eg}   {{\em e.g.}}
\newcommand{\etal} {{\em et al.}}
\newcommand{\ie}   {{\em i.e.}}
\newcommand{\half}  {\frac{1}{2}}
\renewcommand{\bar}{\overline}
\newcommand{\qu}{{\rm q}}
\newcommand{\qb}{${\rm\bar q}$}
\newcommand{\pvec}{\vec p}
\newcommand{\kvec}{\vec k}
\newcommand{\rvec}{\vec r}
\newcommand{\Rvec}{\vec R}
\newcommand{\ieps}{i\varepsilon}
\newcommand{\pl}{{||}}
\newcommand{\order}[1]{${ O}\left(#1 \right)$}
\newcommand{\eq}[1]{(\ref{#1})}
\newcommand{\beq}{\begin{equation}}
\newcommand{\eeq}{\end{equation}}
\newcommand{\M}{{\cal M}}
\newcommand{\VEV}[1]{\left\langle{#1}\right\rangle}
\newcommand{\ket}[1]{\vert\,{#1}\rangle}

\begin{document}

\begin{flushright}
{\small
SLAC--PUB--9056\\
November 2001\\}
\end{flushright}

\vfill

\begin{center}
{{\bf\LARGE QCD Phenomenology and Light-Front\\[1ex] Wavefunctions
}\footnote{Work supported by the Department of Energy under
contract number DE-AC03-76SF00515.}}

\bigskip
Stanley J. Brodsky\\
{\sl Stanford Linear Accelerator Center \\
Stanford University, Stanford, California 94309\\
sjbth@slac.stanford.edu}\\
\medskip
\end{center}

\vfill

\begin{center}
{\it Invited lectures, presented at the\\
   Cracow School Of Theoretical Physics: \\
   41st Course: Fundamental Interactions\\
    Zakopane, Poland\\
    2--11 June 2001
 }\\
\end{center}

\vfill \newpage

\begin{center}
Abstract \end{center}

A natural calculus for describing the bound-state structure of
relativistic composite systems in quantum field theory is the
light-front Fock expansion which encodes the properties of a
hadrons in terms of a set of frame-independent $n-$particle
wavefunctions.  Light-front quantization in the doubly-transverse
light-cone gauge has a number of remarkable advantages, including
explicit unitarity, a physical Fock expansion, the absence of
ghost degrees of freedom, and the decoupling properties needed to
prove factorization theorems in high momentum transfer inclusive
and exclusive reactions.  A number of applications are discussed
in these lectures, including semileptonic $B$ decays, two-photon
exclusive reactions, diffractive dissociation into jets, and
deeply virtual Compton scattering. The relation of the intrinsic
sea to the light-front wavefunctions is  discussed. Light-front
quantization can also be used in the Hamiltonian form to construct
an event generator for high energy physics reactions at the
amplitude level.  The light-cone partition function, summed over
exponentially-weighted light-cone energies, has simple boost
properties which may be useful for studies in heavy ion
collisions.  I also review recent work which shows that the
structure functions measured in deep inelastic lepton scattering
are affected by final-state rescattering, thus modifying their
connection to light-front probability distributions.  In
particular, the shadowing of nuclear structure functions is due to
destructive interference effects from leading-twist diffraction of
the virtual photon, physics not included in the nuclear light-cone
wavefunctions.

\bigskip

\section{Introduction}

 Progress in the development and testing of quantum
chromodynamics will require a detailed understanding of hadron
processes at the amplitude level. For example, exclusive $B$-meson
decays depend critically on the wavefunction of the $B$ as well as
the final-state hadronic wavefunctions. Spin correlations such as
single-spin asymmetries in hard QCD reactions, require an
understanding of the phase structure of hadron amplitudes, physics
well beyond that contained in probability distributions.

One of the challenges of relativistic quantum field theory is to
compute the wavefunctions of bound states, such as the amplitudes
which determine the quark and gluon substructure of hadrons in
quantum chromodynamics.  However, any extension of the
Heisenberg-Schr\"odinger formulation of quantum mechanics $H
\ket{\psi} = i {\partial \over \partial t} \ket{\psi} = E
\ket{\psi}$ to the relativistic domain has to confront seemingly
intractable hurdles: (1) quantum fluctuations preclude finite
particle-number wavefunction representations; (2) the charged
particles arising from the quantum fluctuations of the vacuum
contribute to the matrix element of currents -- thus knowledge of
the wavefunctions alone is insufficient to determine observables;
and (3) the boost of an equal-time wavefunction from one Lorentz
frame to another not only changes particle number, but is as
complicated a dynamical problem as solving for the wavefunction
itself.

In 1949, Dirac~\cite{Dirac:cp} made the remarkable observation
that ordinary ``instant" time $t$ is not the only possible
evolution parameter. In fact,  evolution in ``light-front" time
$\tau = t + z/c = x^+$ has extraordinary advantages for
relativistic systems, stemming from the fact that 7 out of the 10
Poincare' generators, including a Lorentz boost $K_3$, are
kinematical (interaction-independent) when one quantizes a theory
at fixed light-front time.

The light-front fixes the initial boundary conditions of a
composite system as its constituents are intercepted by a
light-wave evaluated on the hyperplane $x^+ = t + z/c$.  In
contrast, determining an atomic wavefunction at a given instant $t
= t_0$ requires measuring the simultaneous scattering of $Z$
photons on the $Z$ electrons. In fact, the Fock-state
representation of bound states defined at equal light-cone time,
\ie, along the light-front, provides wavefunctions of fixed
particle number which are independent of the eigenstate's
four-momentum $P^\mu.$ Furthermore, quantum fluctuations of the
vacuum are absent if one uses light-front time to quantize the
system, so that matrix elements such as the electromagnetic form
factors only depend on the currents of the constituents described
by the light-cone wavefunctions. I will use here the notation
$A^\mu = (A^+, A-, A_\perp),$ where $A^\pm = A^0 \pm A^z$, and the
metric is $A \cdot B = {1\over 2 }( A^+ B^- + A^- B^+ ) - A_\perp
\cdot B_\perp.$

In Dirac's ``Front Form", the generator of light-front time
translations is $P^- = i{ \partial\over \partial \tau}.$ Boundary
conditions are set on the transverse plane labelled by $x_\perp$
and $x^- = z-ct.$ See Fig.~{\ref{fig:bir2}. Given the Lagrangian
of a quantum field theory, $P^-$ can be constructed as an operator
on the Fock basis, the eigenstates of the free theory. Since each
particle in the Fock basis is on its mass shell, $k^- \equiv
k^0-k^3 = {k^2_\perp + m^2 \over k^+},$ and its energy $k^0 =\half
( k^+ + k^-) $ is positive, only particles with positive momenta
$k^+ \equiv k^0 + k^3 \ge 0$ can occur in the Fock basis. Since
the total plus momentum $P^+ = \sum_n k^+_n$ is conserved, the
light-cone vacuum cannot have any particle content.   The operator
$H_{LC} = P^+ P^- - P^2_\perp,$ the ``light-cone Hamiltonian", is
frame-independent.

\begin{figure}
\includegraphics[width=6in]{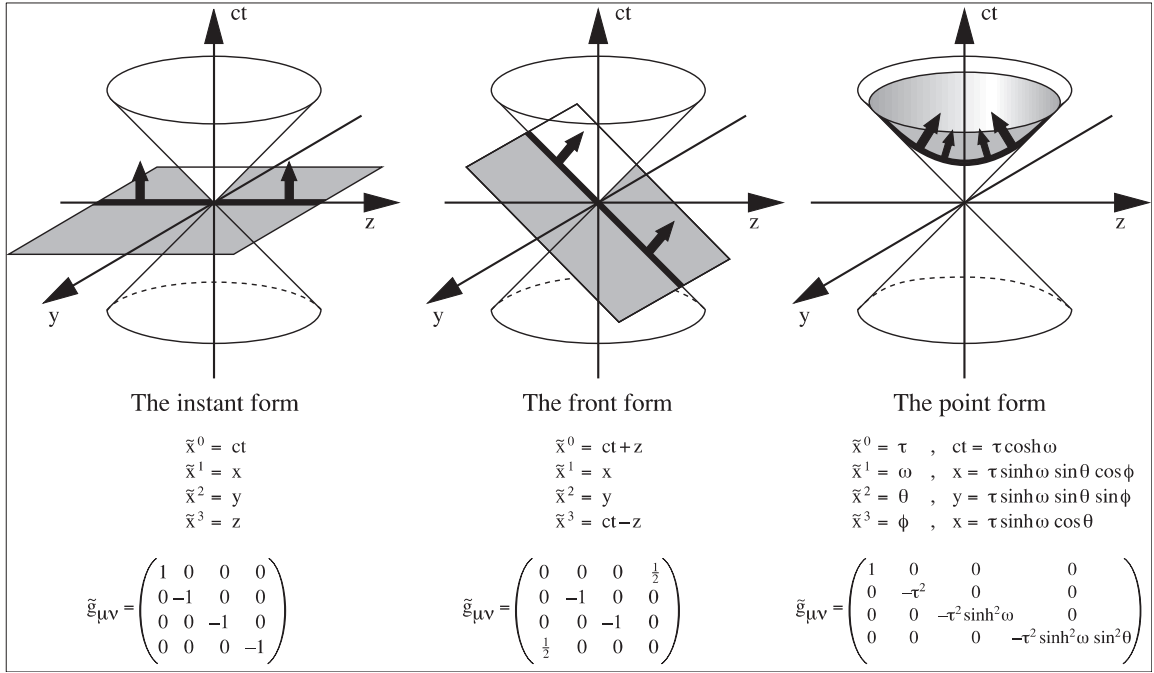}
\caption{\label{fig:bir2}
    Dirac's three forms of Hamiltonian dynamics.
From Ref. \cite{Brodsky:1997de}. }  \end{figure}

The Heisenberg equation on the light-front is
\begin{equation}
H_{LC} \ket{\Psi} = M^2 \ket{\Psi}\ .
\end{equation}
This can in principle be solved by diagonalizing the matrix
$\VEV{n|H_{LC}|m}$ on the free Fock basis:~\cite{Brodsky:1997de}
\begin{equation}
\sum_m \VEV{n|H_{LC}|m}\VEV{m|\psi} = M^2 \VEV{n|\Psi}\ .
\end{equation}
For example the interaction terms of QCD are illustrated in
Fig.~\ref{fig-kyf-3}. The eigenvalues $\{M^2\}$ of
$H_{LC}=H^{0}_{LC} + V_{LC}$ give the squared invariant masses of
the bound and continuum spectrum of the theory. The light-front
Fock space is the eigenstates of the free light-front Hamiltonian;
\ie, it is a Hilbert space of non-interacting quarks and gluons,
each of which satisfy $k^2 = m^2$ and $k^- = {m^2 + k^2_\perp
\over k^+} \ge 0.$  The projections $\{\VEV{n|\Psi}\}$ of the
eigensolution on the $n$-particle Fock states provide the
light-front wavefunctions. Thus solving a quantum field theory is
equivalent to solving a coupled many-body quantum mechanical
problem:
\begin{equation}
\large [M^2 - \sum_{i=1}^n{m_{\perp i}^2\over x_i}\large ] \psi_n
= \sum_{n'}\int \VEV{n|V_{LC}|n'} \psi_{n'}\end{equation}
where the convolution and sum is understood over the Fock number,
transverse momenta, plus momenta, and helicity of the intermediate
states.  Here $m^2_\perp = m^2 + k^2_\perp.$ An essentially
equivalent approach to light-front quantization, pioneered by
Weinberg \cite{Weinberg:1966jm,Brodsky:1973kb}, is to evaluate the
equal-time theory from the perspective of an observer moving in
the negative $\hat z$ direction with arbitrarily large momentum
$P_z \to -\infty.$ The light-cone fraction $x = {k^+\over p^+}$ of
a constituent can be identified with the longitudinal momentum $x
= {k^z\over P^z}$ in a hadron moving with large momentum $P^z.$
Light-front wavefunctions are also related to momentum-space
Bethe-Salpeter wavefunctions by integrating over the relative
momenta $k^- = k^0 - k^z$ since this projects out the dynamics at
$x^+ =0.$

\begin{figure}
\centerline{\includegraphics[height=5in,width=6in]{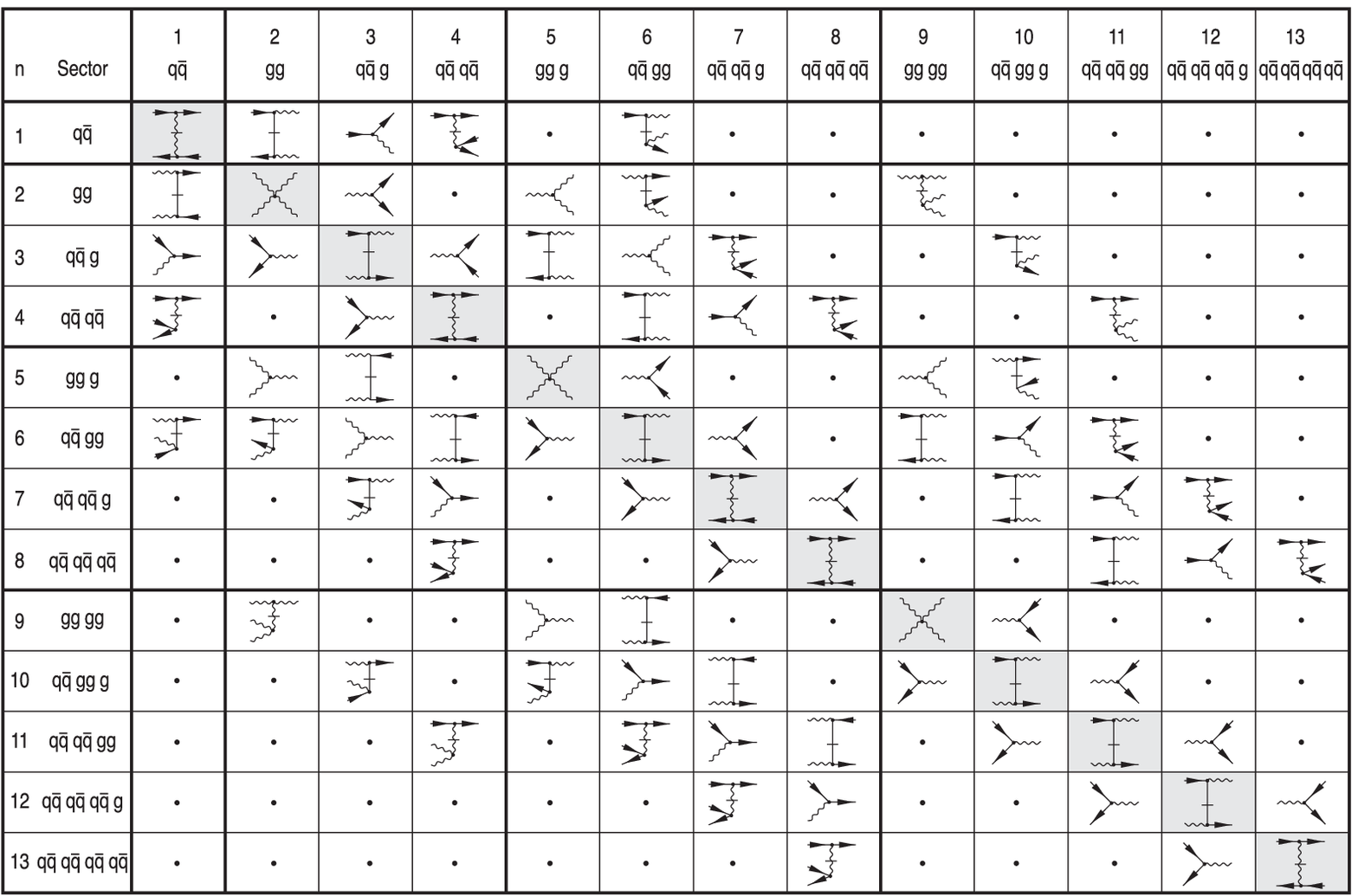}}
\caption{\label{fig-kyf-3} The front-form matrix of QCD
interactions in light-cone gauge. Up to eight constituents in a
meson are shown. From Ref. \cite{Brodsky:1997de} and H. C. Pauli.}
         \end{figure}

We can compare the light-front Fock expansion with the
$n$-particle \break Schr\"odinger momentum space wavefunction
$\psi_N({\vec p}_i)$ of a composite system is the projection of
the exact eigenstate of the equal-time Hamiltonian on the
$n$-particle states of the non-interacting Hamiltonian, the Fock
basis. It represents the amplitude for finding the constituents
with three-momentum ${\vec p}_i$, orbital angular momentum, and
spin, subject to three-momentum conservation and angular momentum
sum rules.  The constituents are on their mass shell, $E_i =
\sqrt{{\vec p}^2_i + m^2_i}$ but do not conserve energy $\sum^n_{i
= 1}E_i > E = \sqrt{{\vec p}^2 + M^2}$.  However, in a
relativistic quantum theory, a bound-state cannot be represented
as a state with a fixed number of constituents. For example, the
existence of gluons which propagate between the valence quarks
necessarily implies that the hadron wavefunction must describe
states with an arbitrary number of gluons.  Thus a hadronic
wavefunction must describe fluctuations in particle number $n$, as
well as momenta and spin.  One has to take into account
fluctuations in the wavefunction which allow for any number of sea
quarks, as long as the total quantum numbers of the constituents
are compatible with the overall quantum numbers of the baryon.

It is especially convenient to develop the light-front formalism
in the light-cone gauge $A^+ = A^0 + A^z = 0$.  In this gauge the
$A^-$ field becomes a dependent degree of freedom, and it can be
eliminated from the gauge theory Hamiltonian, with the addition of
a set of specific instantaneous light-front time interactions.  In
fact in $QCD(1+1)$ theory, this instantaneous interaction provides
the confining linear $x^-$ interaction between quarks.  In $3+1$
dimensions, the transverse field $A^\perp$ propagates massless
spin-one gluon quanta with polarization vectors
\cite{Lepage:1980fj} which satisfy both the gauge condition
$\epsilon^+_\lambda = 0$ and the Lorentz condition $k\cdot
\epsilon= 0$.  Thus no extra condition on the Hilbert space is
required.

In QCD, the wavefunction of a hadron describes its composition in
terms of the momenta and spin projections of quark and gluon
constituents.  For example, the eigensolution of a
negatively-charged meson QCD,  projected on its color-singlet $B =
0$, $Q = -1$, $J_z = 0$ eigenstates $\{\ket{n} \}$ of the free
Hamiltonian $ H^{QCD}_{LC}(g = 0)$ at fixed $\tau = t-z/c$ has the
expansion:
\begin{eqnarray}
\left\vert \Psi_M; P^+, {\vec P_\perp}, \lambda \right> &=&
\sum_{n \ge 2,\lambda_i} \int \Pi^{n}_{i=1} {d^2k_{\perp i} dx_i
\over \sqrt{x_i} 16 \pi^3}
 16 \pi^3 \delta\left(1- \sum^n_j x_j\right) \delta^{(2)}
\left(\sum^n_\ell \vec k_{\perp \ell}\right) \nonumber \\[1ex]
&&\left\vert n; x_i P^+, x_i {\vec P_\perp} + {\vec k_{\perp i}},
\lambda_i\right
> \psi_{n/M}(x_i,{\vec k_{\perp i}},\lambda_i)
 .
\end{eqnarray}
The set of light-front Fock state wavefunctions $\{\psi_{n/M}\}$
represent the ensemble of quark and gluon states possible when the
meson is intercepted at the light-front.  The light-front momentum
fractions $x_i = k^+_i/P^+_\pi = (k^0 + k^z_i)/(P^0+P^z)$ with
$\sum^n_{i=1} x_i = 1$ and ${\vec k_{\perp i}}$ with $\sum^n_{i=1}
{\vec k_{\perp i}} = {\vec 0_\perp}$ represent the relative
momentum coordinates of the QCD constituents and are independent
of the total momentum of the state.

Remarkably, the light-front wavefunctions $\psi_{n/p}(x_i,{\vec
k_{\perp i}},\lambda_i)$ are independent of the proton's momentum
$P^+ = P^0 + P^z$, and $P_\perp$. Thus once one has solved for the
light-front wavefunctions, one can compute hadron matrix elements
of currents between hadronic states of arbitrary momentum. The
actual physical transverse momenta are ${\vec p_{\perp i}} = x_i
{\vec P_\perp} + {\vec k_{\perp i}}.$ The $\lambda_i$ label the
light-front spin $S^z$ projections of the quarks and gluons along
the quantization $z$ direction.  The spinors of the light-front
formalism automatically incorporate the Melosh-Wigner rotation.
The physical gluon polarization vectors $\epsilon^\mu(k,\ \lambda
= \pm 1)$ are specified in light-cone gauge by the conditions $k
\cdot \epsilon = 0,\ \eta \cdot \epsilon = \epsilon^+ = 0.$ The
parton degrees of freedom are thus all physical; there are no
ghost or negative metric states.

The light-front representation thus provides a frame-independent,
quan\-tum-mechan\-ical representation of a hadron at the amplitude
level, capable of encoding its multi-quark, hidden-color and gluon
momentum, helicity, and flavor correlations in the form of
universal process-independent hadron wavefunctions.

Angular momentum has simplifying features in the light-front
formalism since the projection $J_z$ is kinematical and conserved.
Each light-front Fock wavefunction satisfies the angular momentum
sum rule: $ J^z = \sum^n_{i=1} S^z_i + \sum^{n-1}_{j=1} l^z_j \ .
$ The sum over $S^z_i$ represents the contribution of the
intrinsic spins of the $n$ Fock state constituents.  The sum over
orbital angular momenta
\begin{equation}
l^z_j = -{\mathrm i} \left(k^1_j\frac{\partial}{\partial k^2_j}
-k^2_j\frac{\partial}{\partial k^1_j}\right) \end{equation}
 derives from
the $n-1$ relative momenta.  This excludes the contribution to the
orbital angular momentum due to the motion of the center of mass,
which is not an intrinsic property of the hadron. The numerator
structure of the light-front wavefunctions is in large part
determined by the angular momentum constraints.

If one imposes periodic boundary conditions in $x^- = t + z/c$,
then the plus momenta become discrete: $k^+_i = {2\pi \over L}
n_i, P^+ = {2\pi\over L} K$, where $\sum_i n_i = K$
\cite{Maskawa:1975ky,Pauli:1985pv}.  For a given ``harmonic
resolution" $K$, there are only a finite number of ways positive
integers $n_i$ can sum to a positive integer $K$.  Thus at a given
$K$, the dimension of the resulting light-front Fock state
representation of the bound state is rendered finite without
violating Lorentz invariance.  The eigensolutions of a quantum
field theory, both the bound states and continuum solutions, can
then be found by numerically diagonalizing a frame-independent
light-front Hamiltonian $H_{LC}$ on a finite and discrete
momentum-space Fock basis.  Solving a quantum field theory at
fixed light-front time $\tau$ thus can be formulated as a
relativistic extension of Heisenberg's matrix mechanics.  The
continuum limit is reached for $K \to \infty.$ This formulation of
the non-perturbative light-front quantization problem is called
``discretized light-cone quantization" (DLCQ)~\cite{Pauli:1985pv}.
Lattice gauge theory has also been used to calculate the pion
light-front wavefunction~\cite{Abada:2001if}.

The DLCQ method has been used extensively for solving one-space
and one-time theories~\cite{Brodsky:1997de}, including
applications to supersymmetric quantum field
theories~\cite{Matsumura:1995kw} and specific tests of the
Maldacena conjecture~\cite{Hiller:2001mh}. There has been progress
in systematically developing the computation and renormalization
methods needed to make DLCQ viable for QCD in physical spacetime.
For example, John Hiller, Gary McCartor, and
I~\cite{Brodsky:2001ja} have shown how DLCQ can be used to solve
3+1 theories despite the large numbers of degrees of freedom
needed to enumerate the Fock basis.  A key feature of our work is
the introduction of Pauli Villars fields to regulate the UV
divergences and perform renormalization while preserving the
frame-independence of the theory.  A recent application of DLCQ to
a 3+1 quantum field theory with Yukawa interactions is given in
Ref.~\cite{Brodsky:2001ja}.  There has also been important
progress using the transverse lattice, essentially a combination
of DLCQ in 1+1 dimensions together with a lattice in the
transverse
dimensions~\cite{Bardeen:1979xx,Dalley:2001gj,Burkardt:2001dy}.
One can also define a truncated theory by eliminating the higher
Fock states in favor of an effective
potential~\cite{Pauli:2001vi}. Spontaneous symmetry breaking and
other nonperturbative effects associated with the instant-time
vacuum are hidden in dynamical or constrained zero modes on the
light-front.  An introduction is given in
Refs.~\cite{McCartor:hj,Yamawaki:1998cy}.

Because of their Lorentz invariance, it is particularily easy to
write down exact expressions for matrix elements of currents and
other local operators, even the couplings of gravitons. In fact as
I discuss in Section 3, one can show that the anomalous
gravito-magnetic moment $B(0)$, analogous to $F_2(0)$ in
electromagnetic current interactions, vanishes identically for any
system, composite or elementary~\cite{Brodsky:2001ii}. This
important feature which follows in general from the equivalence
principle, is obeyed explicitly in the light-front formalism.

The set of light-front wavefunctions provide a frame-independent,
quantum-mechan\-ical description of hadrons at the amplitude level
capable of encoding multi-quark and gluon momentum, helicity, and
flavor correlations in the form of universal process-independent
hadron wavefunctions. Matrix elements of spacelike currents such
as the spacelike electromagnetic form factors have an exact
representation in terms of simple overlaps of the light-front
wavefunctions in momentum space with the same $x_i$ and unchanged
parton number \cite{Drell:1970km,West:1970av,Brodsky:1980zm}. The
measurement and interpretation of the basic parameters of the
electroweak theory and $CP$ violation depends on an understanding
of the dynamics and phase structure of $B$ decays at the amplitude
level. The light-front Fock representation is specially
advantageous in the study of exclusive $B$ decays. For example, we
can write down an exact frame-independent representation of decay
matrix elements such as $B \to D \ell \bar \nu$ from the overlap
of $n' = n$ parton conserving wavefunctions and the overlap of $n'
= n-2$ from the annihilation of a quark-antiquark pair in the
initial wavefunction \cite{Brodsky:1999hn}.  The handbag
contribution to the leading-twist off-forward parton distributions
measured in deeply virtual Compton scattering have a similar
light-front wavefunction representation as overlap integrals of
light-front wavefunctions \cite{Brodsky:2000xy,Diehl:2000xz}. I
will review this application in Sections 3 and 4.

Factorization theorems have recently been proven which allow one
to rigorously compute certain types of exclusive $B$ decays in
terms of the light-front wavefunctions and distribution amplitudes
of B meson and the final state hadrons.  The proofs are similar to
those used in the analysis of exclusive amplitudes involving large
momentum transfer.  I review this topic in Section 5 and 6.

In principle, the light-front wavefunctions contain fluctuations
of states with arbitrary number of quark and gluon partons. For
example, contains higher Fock states such as $uud s \bar s>$ and
$uud c \bar c>$ which are intrinsic to the physics of the proton
itself; {\em i.e.}, they are multi-connected to the valence quarks
and are not generated by gluon splitting.   A rigorous analysis of
the momentum fraction and spin carried by intrinsic heavy quarks
recently been given by Franz {\em et al}~\cite{Franz:2000ee}.
These quantities scale nominally as $1/m^2_Q$ in non-Abelian gauge
theory, in striking contrast to the $1/m^4_Q$ scaling which
follows from the Euler-Heisenberg Lagrangian in QED.  In general,
the intrinsic sea in the proton is asymmetric between the $Q(x)$
and $\bar Q(x)$ distributions, in contrast to the near symmetry of
quark and antiquark distributions generated by DGLAP evolution.

The fact that the $B$ meson contains Fock states with intrinsic
strangeness and charm leads to a number of new phenomena in
exclusive $B$ decays. In particular, since the charm quarks can
facilitate weak interactions, one can evade the CKM hierarchy.
Susan Gardner and I have shown that the color octet intrinsic
charm Fock components of the $B$ meson can  give significant
modifications of standard predictions for channels such as $B \to
\rho \pi$. I will review this in Section 8.

The quark and gluon probability distributions $q_i(x,Q)$ and
$g(x,Q)$ of a hadron can be computed from the absolute squares of
the light-front wavefunctions, integrated over the transverse
momentum up to the resolution scale $Q$.  All helicity
distributions are thus encoded in terms of the light-front
wavefunctions.  The DGLAP evolution of the structure functions can
be derived from the high $k_\perp$ properties of the light-front
wavefunctions. Thus given the light-front wavefunctions, one can
compute \cite{Lepage:1980fj} all of the leading twist helicity and
transversity distributions measured in polarized deep inelastic
lepton scattering.  For example, the helicity-specific quark
distributions at resolution $\Lambda$ correspond to
\begin{eqnarray}
&&q_{\lambda_q/\Lambda_p}(x, \Lambda) = \sum_{n,q_a}
\int\prod^n_{j=1} {dx_j d^2 k_{\perp j}\over 16 \pi^3}
\sum_{\lambda_i} \vert \psi^{(\Lambda)}_{n/H}(x_i,\vec k_{\perp
i},\lambda_i)\vert^2
\\
&& \qquad\times 16 \pi^3 \delta\left(1- \sum^n_i x_i\right)
\delta^{(2)} \left(\sum^n_i \vec k_{\perp i}\right) \delta(x -
x_q) \delta_{\lambda, \lambda_q} \Theta(\Lambda^2 - {\cal M}^2_n)\
, \nonumber
\end{eqnarray}
where the sum is over all quarks $q_a$ which match the quantum
numbers, light-front momentum fraction $x,$ and helicity of the
struck quark.  Similarly, the transversity distributions and
off-diagonal helicity convolutions are defined as a density matrix
of the light-front wavefunctions.  This defines the LC
factorization scheme \cite{Lepage:1980fj} where the invariant mass
squared ${\cal M}^2_n = \sum_{i = 1}^n {(k_{\perp i}^2 + m_i^2 )/
x_i}$ of the $n$ partons of the light-front wavefunctions is
limited to $ {\cal M}^2_n < \Lambda^2$.

However, it is not true that the leading-twist structure functions
$F_i(x,Q^2)$  measured in deep inelastic lepton scattering are
identical to the quark and gluon distributions. For example, it is
usually assumed, following the parton model,  that the $F_2$
structure function measured in neutral current deep inelastic
lepton scattering is at leading order in $1/Q^2$ simply
$F_2(x,Q^2) =\sum_q  e^2_q  x q(x,Q^2)$, where $x = x_{bj} = Q^2/2
p\cdot q$ and $q(x,Q)$ can be computed from the absolute square of
the proton's light-front wavefunction.   I will report on recent
work by Paul Hoyer, Nils Marchal, Stephane Peigne, Francesco
Sannino, and myself which shows that this standard identification
is wrong. In particular, the shadowing corrections related to the
Gribov-Glauber mechanism, the interference effects of leading
twist diffractive processes in nuclei are separate effects in deep
inelastic scattering, are  not computable from the bound state
wavefunctions of the target nucleon or nucleus.

Remarkably, it is now possible to measure the light-front
wavefunctions of a relativistic hadron by diffractively
dissociating it into jets whose momentum distribution is
correlated with the valence quarks' momenta
\cite{Ashery:1999nq,Bertsch:1981py,Frankfurt:1993it,Frankfurt:2000tq}.
At high energies each light-front Fock state interacts distinctly;
\eg, Fock states with small particle number and small impact
separation have small color dipole moments and can traverse a
nucleus with minimal interactions.  This is the basis for the
predictions for ``color transparency" in hard quasi-exclusive
\cite{Brodsky:1988xz,Frankfurt:1988nt} and diffractive reactions
\cite{Bertsch:1981py,Frankfurt:1993it,Frankfurt:2000tq}. QCD color
transparency thus tests a fundamental ansatz of QCD, that hadronic
interactions are a manifestation of gauge interactions. The E791
experiment  has recently provided a remarkable confirmation of
this consequence of QCD color transparency, a key property of
LCWFs and the gauge field interactions in QCD.  The new EVA
spectrometer experiment E850 at Brookhaven has also reported
striking effects of color transparency in quasi-elastic
proton-proton scattering in nuclei~\cite{Leksanov:2001ui}.  I will
review this important development in Section 7.

The CLEO collaboration has verified the scaling and angular
predictions for hard exclusive two-photon processes such as
$\gamma^* \gamma \to \pi^0$ and $\gamma \gamma \to \pi^+ \pi^-$.
The L3 experiment at LEP at CERN has also measured a number of
exclusive hadron production channels in two-photon processes,
providing important constraints on baryon and meson distribution
amplitudes and checks of perturbative QCD factorization. These
processes are particularly sensitive to the meson distribution
amplitudes, the non-perturbative wavefunctions which control hard
QCD exclusive processes, information essential for progress in
interpreting exclusive $B$ decays.  New data from CLEO (Paar, {\em
et al.}) for $\gamma \gamma \to \pi^+ \pi^+ + K^+ K^-$ at $W =
\sqrt s
> 2. 5$ GeV.  is in striking agreement with the perturbative QCD
prediction given by Lepage and myself.  Moreover, the angular
distribution shows a striking transition to the predicted QCD form
as $W$ is raised.  The $\gamma^* \gamma \to \pi^0$ results are in
close agreement with the scaling and normalization of the PQCD
prediction, provided that the pion distribution amplitude
$\phi_\pi(x,Q)$ is close to the $x(1-x)$ form, the asymptotic
solution to the evolution equation.  In Section 6 I review the
theory and emphasized the need for more such meson pair production
data, particularly measurements of ratios and angular dependencies
which are particularly sensitive to the meson and baryon
distribution amplitudes \cite{Lepage:1980fj}, $\phi_M(x,Q)$, and
$\phi_B(x_i,Q)$. These quantities specify how a hadron shares its
longitudinal momentum among its valence quarks; they control
virtually all exclusive processes involving a hard scale $Q$,
including form factors, Compton scattering and photoproduction at
large momentum transfer, as well as the decay of a heavy hadron
into specific final states \cite{Beneke:1999br,Keum:2000ph}.

The discretized light-front quantization method developed by H.C.
Pauli and myself~\cite{Pauli:1985ps} is a powerful technique for
finding the non-perturbative solutions of quantum field theories.
The basic method is to diagonalize the light-front Hamiltonian in
a light-front Fock basis defined using periodic boundary
conditions in $x^-$ and $x_\perp$. The method preserves the
frame-independence of the Front form. The DLCQ method is now used
extensively to solve one-space and one-time theories, including
supersymmetric theories. New applications of DLCQ to
supersymmetric quantum field theories and specific tests of the
Maldacena conjecture  have recently been given by Pinsky and
Trittman.

There has been progress recently in systematically developing the
computation and renormalization methods needed to make DLCQ viable
for QCD in physical spacetime. Recently John Hiller, Gary McCartor
and I have shown how DLCQ can be used to solve 3+1 theories
despite the large numbers of degrees of freedom needed to
enumerate the Fock basis~\cite{Brodsky:2001ja}.   A key feature of
our work, is the introduction of Pauli Villars fields in order to
regulate the UV divergences and perform renormalization, again
while preserving the frame-independence of the theory.  Further
discussion will be given in Section 9.  A review of DLCQ and its
applications  is given in Ref. \cite{Brodsky:1998de}.  There also
has been important progress using the transverse lattice,
essentially a combination of DLCQ in i+1 dimensions together with
a lattice in the transverse space.

Models of the light-front wavefunction are important in the
absence of exact solutions. A simple but potentially useful model
developed by Dae Sung Hwang and myself is discussed in Section 10.

The interaction Hamiltonian of QCD in light-cone gauge can be
derived by systematically applying the Dirac bracket method to
identify the independent fields \cite{Srivastava:2000cf}. It
contains the usual Dirac interactions between the quarks and
gluons, the three-point and four-point gluon non-Abelian
interactions plus instantaneous light-front-time gluon exchange
and quark exchange contributions
\begin{eqnarray}
{\cal H}_{int}&=&
  -g \,{{\bar\psi}}^{i}
\gamma^{\mu}{A_{\mu}}^{ij}{{\psi}}^{j}   \nonumber \\
&& +\frac{g}{2}\, f^{abc} \,(\partial_{\mu}{A^{a}}_{\nu}-
\partial_{\nu}{A^{a}}_{\mu}) A^{b\mu} A^{c\nu} \nonumber \\
&& +\frac {g^2}{4}\,
f^{abc}f^{ade} {A_{b\mu}} {A^{d\mu}} A_{c\nu} A^{e\nu} \nonumber \\
&& - \frac{g^{2}}{ 2}\,\, {{\bar\psi}}^{i} \gamma^{+}
\,(\gamma^{\perp'}{A_{\perp'}})^{ij}\,\frac{1}{i\partial_{-}} \,
(\gamma^{\perp} {A_{\perp}})^{jk}\,{\psi}^{k} \nonumber \\
&& -\frac{g^{2}}{ 2}\,{j^{+}}_{a}\, \frac
{1}{(\partial_{-})^{2}}\, {j^{+}}_{a}
\end{eqnarray}
where
\begin{equation}
{j^{+}}_{a}={{\bar\psi}}^{i} \gamma^{+} (
{t_{a}})^{ij}{{\psi}}^{j} + f_{abc} (\partial_{-} A_{b\mu})
A^{c\mu} \ .
\end{equation}

In light-front time-ordered perturbation theory, a Green's
functions is expanded as a power series in the interactions with
light-front energy denominators $\sum_{\rm initial} k^-_i -
\sum_{\rm intermediate} k^-_i + i \epsilon$ replacing the usual
energy denominators.  [For a review see Ref.
\cite{Brodsky:1989pv}.] In general each Feynman diagram with $n$
vertices corresponds to the sum of $n!$ time-ordered
contributions.  However, in light-front-time-ordered perturbation
theory, only those few graphs where all $k^+_i \ge 0$ survive.  In
addition the form of the light-front kinetic energies is rational:
$k^- = {k^2_\perp + m^2 \over k^+}$, replacing the nonanalytic
$k^0 = \sqrt{{\vec k}^2 + m^2}$ of equal-time theory. Thus
light-front-time-ordered perturbation theory provides a viable
computational method where one can trace the physical evolution of
a process.  The integration measures are only three-dimensional
$d^2k_\perp dx$; in effect, the $k^-$ integral of the covariant
perturbation theory is performed automatically.

Alternatively, one derive Feynman rules for QCD in light-cone
gauge, thus allowing the use of standard covariant computational
tools and renormalization methods including dimensional
regularization.  Prem Srivastava and I \cite{Srivastava:2000cf}
have recently presented a systematic study of
light-front-quantized gauge theory in light-cone gauge using a
Dyson-Wick S-matrix expansion based on light-front-time-ordered
products.  The gluon propagator has the form
\begin{equation}
\VEV{0|\,T({A^{a}}_{\mu}(x){A^{b}}_{\nu}(0))\,|0} ={{i\delta^{ab}}
\over {(2\pi)^{4}}} \int d^{4}k \;e^{-ik\cdot x} \; \;
{D_{\mu\nu}(k)\over {k^{2}+i\epsilon}}
\end{equation}
where we have defined
\begin{equation}
D_{\mu\nu}(k)= D_{\nu\mu}(k)= -g_{\mu\nu} + \frac
{n_{\mu}k_{\nu}+n_{\nu}k_{\mu}}{(n\cdot k)} - \frac {k^{2}}
{(n\cdot k)^{2}} \, n_{\mu}n_{\nu}.
\end{equation}
Here $n_{\mu}$ is a null four-vector, gauge direction, whose
components are chosen to be $\, n_{\mu}={\delta_{\mu}}^{+}$, $\,
n^{\mu}={\delta^{\mu}}_{-}$.  Note also
\begin{eqnarray}
D_{\mu\lambda}(k) {D^{\lambda}}_{\nu}(k)=
D_{\mu\perp}(k) {D^{\perp}}_{\nu}(k)&=& - D_{\mu\nu}(k), \nonumber \\
k^{\mu}D_{\mu\nu}(k)=0, \qquad \quad && n^{\mu}D_{\mu\nu}(k)\equiv
D_{-\nu}(k)=0, \nonumber \\ D_{\lambda\mu}(q) \,D^{\mu\nu}(k)\,
D_{\nu\rho}(q') &=& -D_{\lambda\mu}(q)D^{\mu\rho}(q').
\end{eqnarray}
The gauge field propagator $\,\,i\,D_{\mu\nu}(k)/
(k^{2}+i\epsilon)\,$ is transverse not only to the gauge direction
$n_{\mu}$ but also to $k_{\mu}$, {\em i.e.}, it is {\it
doubly-transverse}.  This leads to appreciable simplifications in
the computations in QCD. For example, the coupling of gluons to
propagators carrying high momenta is automatic.  The absence of
collinear divergences in irreducible diagrams in the light-cone
gauge greatly simplifies the leading-twist factorization of soft
and hard gluonic corrections in high momentum transfer inclusive
and exclusive reactions \cite{Lepage:1980fj} since the numerators
associated with the gluon coupling only have transverse
components. The renormalization factors in the light-cone gauge
are independent of the reference direction $n^\mu$.  Since the
gluon only has physical polarization, its renormalization factors
satisfy $Z_1=Z_3$.  Because of its explicit unitarity in each
graph, the doubly-transverse gauge is well suited for calculations
identifying the
 ``pinch" effective charge \cite{Cornwall:1989gv,Brodsky:2000cr}.

The running coupling constant and QCD $\beta$ function have also
been computed at one loop in the doubly-transverse light-cone
gauge \cite{Srivastava:2000cf}. It is also possible to effectively
quantize QCD using light-front methods in covariant Feynman gauge
\cite{Srivastava:2000gi}.

A remarkable advantage of light-front quantization is that the
vacuum state $\ket{0}$ of the full QCD Hamiltonian evidently
coincides with the free vacuum.  The light-front vacuum is
effectively trivial if the interaction Hamiltonian applied to the
perturbative vacuum is zero. Note that all particles in the
Hilbert space have positive energy $k^0 = {1\over 2}(k^+ + k^-)$,
and thus positive light-front $k^\pm$.  Since the plus momenta
$\sum k^+_i$ is conserved by the interactions, the perturbative
vacuum can only couple to states with particles in which all
$k^+_i$ = 0; \ie, so called zero-mode states.  In the case of QED,
a massive electron cannot have $k^+ = 0$ unless it also has
infinite energy.  In a remarkable calculation, Bassetto and
collaborators \cite{Bassetto:1999tm} have shown that the
computation of the spectrum of $QCD(1+1)$ in equal time
quantization requires constructing the full spectrum of non
perturbative contributions (instantons).  In contrast, in the
light-front quantization of gauge theory, where the $k^+ = 0 $
singularity of the instantaneous interaction is defined by a
simple infrared regularization, one obtains the correct spectrum
of $QCD(1+1)$ without any need for vacuum-related contributions.

In the case of QCD(3+1), the momentum-independent four-gluon
non-Abelian interaction in principle can couple the perturbative
vacuum to a state with four collinear gluons in which all of the
gluons have all components $k^\mu_i = 0,$ thus hinting at role for
zero modes in theories with massless quanta.  In fact, zero modes
of auxiliary fields are necessary to distinguish the theta-vacua
of massless QED(1+1)
\cite{Yamawaki:1998cy,McCartor:2000yy,Srivastava:1999et}, or to
represent a theory in the presence of static external boundary
conditions or other constraints.  Zero-modes provide the
light-front representation of spontaneous symmetry breaking in
scalar theories \cite{Pinsky:1994yi}.

There are other applications of the light-front formalism:

1.  The distribution of spectator particles in the final state in
the proton fragmentation region in deep inelastic scattering at an
electron-proton collider are encoded in the light-front
wavefunctions of the target proton.  Conversely, the light-front
wavefunctions can be used to describe the coalescence of comoving
quarks into final state hadrons.

2.  The light-front wavefunctions also specify the multi-quark and
gluon correlations of the hadron.  Despite the many sources of
power-law corrections to the deep inelastic cross section, certain
types of dynamical contributions will stand out at large $x_{bj}$
since they arise from compact, highly-correlated fluctuations of
the proton wavefunction.  In particular, there are particularly
interesting dynamical ${\cal O}(1/Q^2)$ corrections which are due
to the {\it interference} of quark currents; {\it i.e.},
contributions which involve leptons scattering amplitudes from two
different quarks of the target nucleon \cite{Brodsky:2000zu}.

3.  The higher Fock states of the light hadrons describe the sea
quark structure of the deep inelastic structure functions,
including ``intrinsic" strange\-ness and charm fluctuations
specific to the hadron's structure rather than gluon substructure
\cite{Brodsky:1980pb,Harris:1996jx}. Ladder relations connecting
state of different particle number follow from the QCD equation of
motion and lead to Regge behavior of the quark and gluon
distributions at $x \to 0$ \cite{Antonuccio:1997tw}.

4.  The gauge- and process-independent meson and baryon
valence-quark distribution amplitudes $\phi_M(x,Q)$, and
$\phi_B(x_i,Q)$ which control exclusive processes involving a hard
scale $Q$, including heavy quark decays, are given by the valence
light-front Fock state wavefunctions integrated over the
transverse momentum up to the resolution scale $Q$.  The evolution
equations for distribution amplitudes follow from the perturbative
high transverse momentum behavior of the light-front wavefunctions
\cite{Brodsky:1989pv}.

5.  Proofs of factorization theorems in hard exclusive and
inclusive reactions are greatly simplified since the propagating
gluons in light-cone gauge couple only to transverse currents;
collinear divergences are thus automatically suppressed.

6. The deuteron form factor at high $Q^2$ is sensitive to
wavefunction configurations where all six quarks overlap within an
impact separation $b_{\perp i} < {\cal O} (1/Q).$ The leading
power-law fall off predicted by QCD is $F_d(Q^2) =
f(\alpha_s(Q^2))/(Q^2)^5$, where, asymptotically,
$f(\alpha_s(Q^2)) \propto \alpha_s(Q^2)^{5+2\gamma}$
\cite{Brodsky:1976rz,Brodsky:1983vf}. In general, the six-quark
wavefunction of a deuteron is a mixture of five different
color-singlet states.  The dominant color configuration at large
distances corresponds to the usual proton-neutron bound state.
However at small impact space separation, all five Fock
color-singlet components eventually evolve to a state with equal
weight, \ie, the deuteron wavefunction evolves to 80\%\ ``hidden
color'' \cite{Brodsky:1983vf}. The relatively large normalization
of the deuteron form factor observed at large $Q^2$ hints at
sizable hidden-color contributions \cite{Farrar:1991qi}. Hidden
color components can also play a predominant role in the reaction
$\gamma d \to J/\psi p n$ at threshold if it is dominated by the
multi-fusion process $\gamma g g \to J/\psi$
\cite{Brodsky:2000zc}.  Hard exclusive nuclear processes can also
be analyzed in terms of ``reduced amplitudes" which remove the
effects of nucleon substructure.

Light-front wavefunctions are thus the frame-independent
interpolating functions between hadron and quark and gluon degrees
of freedom.  Hadron amplitudes are computed from the convolution
of the light-front wavefunctions with irreducible quark-gluon
amplitudes.  More generally, all multi-quark and gluon
correlations in the bound state are represented by the light-front
wavefunctions. The light-front Fock representation is thus a
representation of the underlying quantum field theory.  I will
discuss progress in computing light-front wavefunctions directly
from QCD in Sections 9 and 10.

Light-front quantization can also be used in the Hamiltonian form
to construct an event generator for high energy physics reactions
at the amplitude level.  The light-front partition function,
summed over exponentially-weighted light-front energies, has
simple boost properties which may be useful for studies in heavy
ion collisions. I discuss these topics in Sections 14 and 15.

\section{Other Theoretical Tools}

In addition to the light-front Fock expansion, a number of other
useful theoretical tools are available to eliminate theoretical
ambiguities in QCD predictions:

(1) Conformal symmetry provides a template for QCD predictions
\cite{Brodsky:1999gm}, leading to relations between observables
which are present even in a theory which is not scale invariant.
For example, the natural representation of distribution amplitudes
is in terms of an expansion of orthonormal conformal functions
multiplied by anomalous dimensions determined by QCD evolution
equations \cite{Brodsky:1980ny,Muller:1994hg,Braun:1999te}. Thus
an important guide in QCD analyses is to identify the underlying
conformal relations of QCD which are manifest if we drop quark
masses and effects due to the running of the QCD couplings.  In
fact, if QCD has an infrared fixed point (vanishing of the
Gell-Mann-Low function at low momenta), the theory will closely
resemble a scale-free conformally symmetric theory in many
applications.

(2) Commensurate scale relations
\cite{Brodsky:1995eh,Brodsky:1998ua} are perturbative QCD
predictions which relate observable to observable at fixed
relative scale, such as the ``generalized Crewther relation"
\cite{Brodsky:1996tb}, which connects the Bjorken and
Gross-Llewellyn Smith deep inelastic scattering sum rules to
measurements of the $e^+ e^-$ annihilation cross section.  Such
relations have no renormalization scale or scheme ambiguity.  The
coefficients in the perturbative series for commensurate scale
relations are identical to those of conformal QCD; thus no
infrared renormalons are present \cite{Brodsky:1999gm}. One can
identify the required conformal coefficients at any finite order
by expanding the coefficients of the usual PQCD expansion around a
formal infrared fixed point, as in the Banks-Zak method
\cite{Brodsky:2000cr}. All non-conformal effects are absorbed by
fixing the ratio of the respective momentum transfer and energy
scales.  In the case of fixed-point theories, commensurate scale
relations relate both the ratio of couplings and the ratio of
scales as the fixed point is approached
 \cite{Brodsky:1999gm}.

(3) $\alpha_V$ and Skeleton Schemes.  A physically natural scheme
for defining the QCD coupling in exclusive and other processes is
the $\alpha_V(Q^2)$ scheme defined from the potential of static
heavy quarks.  Heavy-quark lattice gauge theory can provide highly
precise values for the coupling.  All vacuum polarization
corrections due to fermion pairs are then automatically and
analytically incorporated into the Gell Mann-Low function, thus
avoiding the problem of explicitly computing and resumming quark
mass corrections related to the running of the coupling
\cite{Brodsky:1998mf}. The use of a finite effective charge such
as $\alpha_V$ as the expansion parameter also provides a basis for
regulating the infrared nonperturbative domain of the QCD
coupling.  A similar coupling and scheme can be based on an
assumed skeleton expansion of the theory
\cite{Cornwall:1989gv,Brodsky:2000cr}.

(4) The Abelian Correspondence Principle.  One can consider QCD
predictions as analytic functions of the number of colors $N_C$
and flavors $N_F$.  In particular, one can show at all orders of
perturbation theory that PQCD predictions reduce to those of an
Abelian theory at $N_C \to 0$ with ${\widehat \alpha} = C_F
\alpha_s$ and ${\widehat N_F} = 2 N_F/C_F$ held fixed
\cite{Brodsky:1997jk}. There is thus a deep connection between QCD
processes and their corresponding QED analogs.

\section{Applications of Light-Front Wavefunctions to
\hfill\break Current Matrix Elements}

The light-front Fock representation of current matrix elements has
a number of simplifying properties. The space-like local operators
for the coupling of photons, gravitons and the deep inelastic
structure functions can all be expressed as overlaps of
light-front wavefunctions with the same number of Fock
constituents. This is possible since one can choose the special
frame $q^+ = 0$ \cite{Drell:1970km,West:1970av} for space-like
momentum transfer and take matrix elements of ``plus" components
of currents such as $J^+$ and $T^{++}$.  No contributions to the
current matrix elements from vacuum fluctuations occur.
Similarly, given the local operators for the energy-momentum
tensor $T^{\mu \nu}(x)$ and the angular momentum tensor $M^{\mu
\nu \lambda}(x)$, one can directly compute momentum fractions,
spin properties, and the form factors $A(q^2)$ and $B(q^2)$
appearing in the coupling of gravitons to composite systems
\cite{Brodsky:2001ii}.

In the case of a spin-${1\over 2}$ composite system, the Dirac and
Pauli form factors $F_1(q^2)$ and $F_2(q^2)$ are defined by
\begin{equation}
    \langle P'| J^\mu (0) |P\rangle
       = \bar u(P')\, \Big[\, F_1(q^2)\gamma^\mu +
F_2(q^2){i\over 2M}\sigma^{\mu\alpha}q_\alpha\, \Big] \, u(P)\ ,
\label{Drell1}
\end{equation}
where $q^\mu = (P' -P)^\mu$ and $u(P)$ is the bound state spinor.
In the light-front formalism it is convenient to identify the
Dirac and Pauli form factors from the helicity-conserving and
helicity-flip vector current matrix elements of the $J^+$ current
\cite{Brodsky:1980zm}:
\begin{equation}
\VEV{P+q,\uparrow\left|\frac{J^+(0)}{2P^+} \right|P,\uparrow}
=F_1(q^2) \ , \label{BD1}
\end{equation}
\begin{equation}
\VEV{P+q,\uparrow\left|\frac{J^+(0)}{2P^+}\right|P,\downarrow}
=-(q^1-{\mathrm i} q^2){F_2(q^2)\over 2M}\ . \label{BD2}
\end{equation}
The magnetic moment of a composite system is one of its most basic
properties.  The magnetic moment is defined at the $q^2 \to 0$
limit,
\begin{equation}
\mu=\frac{e}{2 M}\left[ F_1(0)+F_2(0) \right] , \label{DPmu}
\end{equation}
where $e$ is the charge and $M$ is the mass of the composite
system.  We use the standard light-front frame ($q^{\pm}=q^0\pm
q^3$):
\begin{eqnarray}
q &=& (q^+,q^-,{\vec q}_{\perp}) = \left(0, \frac{-q^2}{P^+},
{\vec q}_{\perp}\right), \nonumber \\
P &=& (P^+,P^-,{\vec P}_{\perp}) = \left(P^+, \frac{M^2}{P^+},
{\vec 0}_{\perp}\right), \label{LCF}
\end{eqnarray}
where $q^2=-2 P \cdot q= -{\vec q}_{\perp}^2$ is 4-momentum square
transferred by the photon.

The Pauli form factor and the anomalous magnetic moment $\kappa =
{e\over 2 M} F_2(0)$ can then be calculated from the expression
\begin{equation}
-(q^1-{\mathrm i} q^2){F_2(q^2)\over 2M} = \sum_a  \int {{\mathrm
d}^2 {\vec k}_{\perp} {\mathrm d} x \over 16 \pi^3} \sum_j e_j \
\psi^{\uparrow *}_{a}(x_i,{\vec k}^\prime_{\perp i},\lambda_i) \,
\psi^\downarrow_{a} (x_i, {\vec k}_{\perp i},\lambda_i) {}\ ,
\label{LCmu}
\end{equation}
where the summation is over all contributing Fock states $a$ and
struck constituent charges $e_j$.  The arguments of the
final-state light-front wavefunction are
\begin{equation}
{\vec k}'_{\perp i}={\vec k}_{\perp i}+(1-x_i){\vec q}_{\perp}
\label{kprime1}
\end{equation}
for the struck constituent and
\begin{equation}
{\vec k}'_{\perp i}={\vec k}_{\perp i}-x_i{\vec q}_{\perp}
\label{kprime2}
\end{equation}
for each spectator. Notice that the magnetic moment must be
calculated from the spin-flip non-forward matrix element of the
current. In the ultra-relativistic limit where the radius of the
system is small compared to its Compton scale $1/M$, the anomalous
magnetic moment must vanish \cite{Brodsky:1980zm}. The light-front
formalism is consistent with this theorem.

The form factors of the energy-momentum tensor for a spin-$\half$
\ composite are defined by
\begin{eqnarray}
      \langle P'| T^{\mu\nu} (0)|P \rangle
       &=& \bar u(P')\, \Big[\, A(q^2)
       \gamma^{(\mu} \bar P^{\nu)} +
   B(q^2){i\over 2M} \bar P^{(\mu} \sigma^{\nu)\alpha}
q_\alpha \nonumber \\
   &&\qquad\qquad +  C(q^2){1\over M}(q^\mu q^\nu - g^{\mu\nu}q^2)
    \, \Big]\, u(P) \ ,
\label{Ji12}
\end{eqnarray}
where $q^\mu = (P'-P)^\mu$, $\bar P^\mu={1\over 2}(P'+P)^\mu$,
$a^{(\mu}b^{\nu)}={1\over 2}(a^\mu b^\nu +a^\nu b^\mu)$, and
$u(P)$ is the spinor of the system. One can also readily obtain
the light-front representation of the $A(q^2)$ and $B(q^2)$ form
factors of the energy-tensor Eq. (\ref{Ji12})
\cite{Brodsky:2001ii}. In the interaction picture, only the
non-interacting parts of the energy momentum tensor $T^{+ +}(0)$
need to be computed:
\begin{equation}
\VEV{P+q,\uparrow\left|\frac{T^{++}(0)}{2(P^+)^2}
\right|P,\uparrow} =A(q^2)\ , \label{eBD1}
\end{equation}
\begin{equation}
\VEV{P+q,\uparrow\left|\frac{T^{++}(0)}{2(P^+)^2}\right|P,\downarrow}=
-(q^1-{\mathrm i} q^2){B(q^2)\over 2M}\ . \label{eBD2}
\end{equation}
The $A(q^2)$ and $B(q^2)$ form factors Eqs. (\ref{eBD1}) and
(\ref{eBD2}) are similar to the $F_1(q^2)$ and $F_2(q^2)$ form
factors Eqs.  (\ref{BD1}) and (\ref{BD2}) with an additional
factor of the light-front momentum fraction $x=k^+/P^+$ of the
struck constituent in the integrand.  The $B(q^2)$ form factor is
obtained from the non-forward spin-flip amplitude.  The value of
$B(0)$ is obtained in the $q^2 \to 0$ limit. The angular momentum
projection of a state is given by
\begin{eqnarray}
\VEV{J^i} &=& {1\over 2} \epsilon^{i j k} \int d^3x \VEV{T^{0
k}x^j - T^{0 j} x^k}\nonumber \\[1ex] &=& A(0) \VEV{L^i} + \left[A(0) +
B(0)\right] \bar u(P){1\over 2}\sigma^i u(P) \ . \label{Ji13a}
\end{eqnarray}
This result is derived using a wave-packet description of the
state.  The $\VEV{L^i}$ term is the orbital angular momentum of
the center of mass motion with respect to an arbitrary origin and
can be dropped.  The coefficient of the $\VEV{L^i}$ term must be
1; $A(0) = 1 $ also follows when we evaluate the four-momentum
expectation value $\VEV{P^\mu}$. Thus the total intrinsic angular
momentum $J^z$ of a nucleon can be identified with the values of
the form factors $A(q^2)$ and $B(q^2)$ at $q^2= 0:$
\begin{equation}
      \VEV{J^z} = \VEV{{1\over 2} \sigma^z} \left[A(0) + B(0)\right] \ .
\label{Ji13}
\end{equation}

The anomalous moment coupling $B(0)$ to a graviton can in fact be
shown to vanish for any composite system.  This remarkable result,
first derived by Okun and Kobzarev
\cite{Okun,Ji:1996kb,Ji:1997ek,Ji:1997nm,Teryaev:1999su}, follows
directly from the Lorentz boost properties of the light-front Fock
representation \cite{Brodsky:2001ii}.

Dae Sung Hwang, Bo-Qiang Ma, Ivan Schmidt, and I
\cite{Brodsky:2001ii} have recently shown that the light-front
wavefunctions generated by the radiative corrections to the
electron in QED provides a simple system for understanding the
spin and angular momentum decomposition of relativistic systems.
This perturbative model also illustrates the interconnections
between Fock states of different number.  The model is patterned
after the quantum structure which occurs in the one-loop Schwinger
${\alpha / 2 \pi} $ correction to the electron magnetic moment
\cite{Brodsky:1980zm}. In effect, we can represent a spin-$\half$
~ system as a composite of a spin-$\half$ ~ fermion and spin-one
vector boson with arbitrary masses.  A similar model has been used
to illustrate the matrix elements and evolution of light-front
helicity and orbital angular momentum operators
\cite{Harindranath:1999ve}. This representation of a composite
system is particularly useful because it is based on two
constituents but yet is totally relativistic.  We can then
explicitly compute the form factors $F_1(q^2)$ and $F_2(q^2)$ of
the electromagnetic current, and the various contributions to the
form factors $A(q^2)$ and $B(q^2)$ of the energy-momentum tensor.

\begin{figure}
\begin{center}
\includegraphics[height=4in]{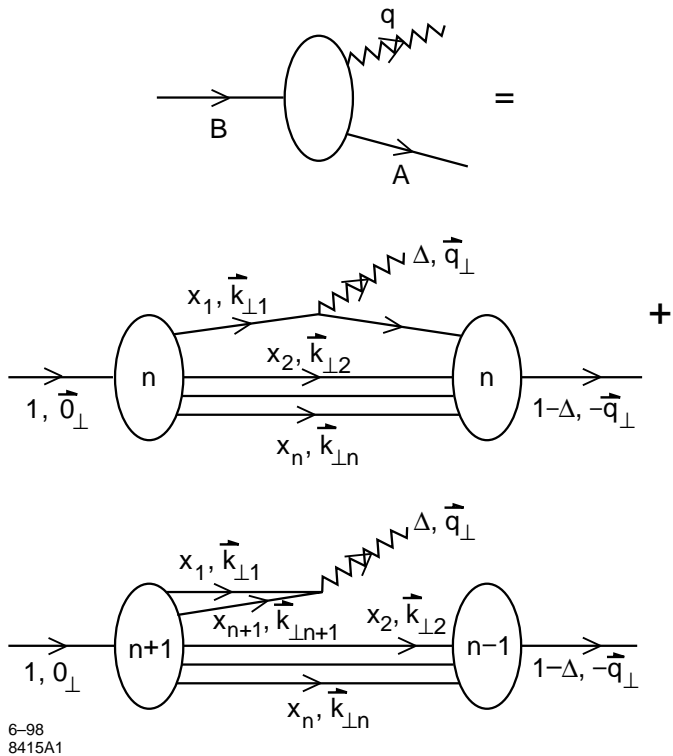}
\end{center}
\caption[*]{Exact representation of electroweak decays and
time-like form factors in the light-front Fock representation.
\label{fig1}}
\end{figure}

Another remarkable advantage of the light-front formalism is that
exclusive semileptonic $B$-decay amplitudes such as $B\rightarrow
A \ell \bar{\nu}$ can also be evaluated exactly
\cite{Brodsky:1999hn}. The time-like decay matrix elements require
the computation of the diagonal matrix element $n \rightarrow n$
where parton number is conserved, and the off-diagonal
$n+1\rightarrow n-1$ convolution where the current operator
annihilates a $q{\bar{q'}}$ pair in the initial $B$ wavefunction.
See Fig.  \ref{fig1}.  This term is a consequence of the fact that
the time-like decay $q^2 = (p_\ell + p_{\bar{\nu}} )^2 > 0$
requires a positive light-front momentum fraction $q^+ > 0$.
Conversely for space-like currents, one can choose $q^+=0$, as in
the Drell-Yan-West representation of the space-like
electromagnetic form factors.  However, as can be seen from the
explicit analysis of the form factor in a perturbative model, the
off-diagonal convolution can yield a nonzero $q^+/q^+$ limiting
form as $q^+ \rightarrow 0$.  This extra term appears specifically
in the case of ``bad" currents such as $J^-$ in which the coupling
to $q\bar q$ fluctuations in the light-front wavefunctions are
favored.  In effect, the $q^+ \rightarrow 0$ limit generates
$\delta(x)$ contributions as residues of the $n+1\rightarrow n-1$
contributions.  The necessity for such ``zero mode" $\delta(x)$
terms has been noted by Chang, Root and Yan \cite{Chang:1973xt},
Burkardt \cite{Burkardt:1989wy}, and Ji and Choi
\cite{Choi:1998nf}.

The off-diagonal $n+1 \rightarrow n-1$ contributions give a new
perspective for the physics of $B$-decays.  A semileptonic decay
involves not only matrix elements where a quark changes flavor,
but also a contribution where the leptonic pair is created from
the annihilation of a $q {\bar{q'}}$ pair within the Fock states
of the initial $B$ wavefunction.  The semileptonic decay thus can
occur from the annihilation of a nonvalence quark-antiquark pair
in the initial hadron. This feature will carry over to exclusive
hadronic $B$-decays, such as $B^0 \rightarrow \pi^-D^+$.  In this
case the pion can be produced from the coalescence of a $d\bar u$
pair emerging from the initial higher particle number Fock
wavefunction of the $B$.  The $D$ meson is then formed from the
remaining quarks after the internal exchange of a $W$ boson.

In principle, a precise evaluation of the hadronic matrix elements
needed for $B$-decays and other exclusive electroweak decay
amplitudes requires knowledge of all of the light-front Fock
wavefunctions of the initial and final state hadrons.  In the case
of model gauge theories such as QCD(1+1) \cite{Hornbostel:1990fb}
or collinear QCD \cite{Antonuccio:1995fs} in one-space and
one-time dimensions, the complete evaluation of the light-front
wavefunction is possible for each baryon or meson bound-state
using the DLCQ method.  It would be interesting to use such
solutions as a model for physical $B$-decays.

\section{Light-front Representation of Deeply Virtual
\hfill\break Compton Scattering}

The virtual Compton scattering process ${d\sigma\over dt}(\gamma^*
p \to \gamma p)$ for large initial photon virtuality $q^2=-Q^2$
has extraordinary sensitivity to fundamental features of the
proton's structure.  Even though the final state photon is
on-shell, the deeply virtual process probes the elementary quark
structure of the proton near the light front as an effective local
current.  In contrast to deep inelastic scattering, which measures
only the absorptive part of the forward virtual Compton amplitude
$Im {\cal T}_{\gamma^* p \to \gamma^* p}$, deeply virtual Compton
scattering allows the measurement of the phase and spin structure
of proton matrix elements for general momentum transfer squared
$t$.  In addition, the interference of the virtual Compton
amplitude and Bethe-Heitler wide angle scattering Bremsstrahlung
amplitude where the photon is emitted from the lepton line leads
to an electron-positron asymmetry in the ${e^\pm  p \to e^\pm
\gamma p}$ cross section which is proportional to the real part of
the Compton amplitude
\cite{Brodsky:1972zh,Brodsky:1972vv,Brodsky:1973hm}.  The deeply
virtual Compton amplitude $\gamma^* p \to \gamma p$ is related by
crossing to another important process $\gamma^* \gamma \to $
hadron pairs at fixed invariant mass which can be measured in
electron-photon collisions \cite{Diehl:2000uv}.

To leading order in $1/Q$, the deeply virtual Compton scattering
amplitude \hfill\break $\gamma^*(q) p(P) \to \gamma(q') p(P')$
factorizes as the convolution in $x$ of the amplitude $t^{\mu
\nu}$ for hard Compton scattering on a quark line with the
generalized Compton form factors $H(x,t,\zeta),$ $ E(x,t,\zeta)$,
$\tilde H(x,t,\zeta),$ and $\tilde E(x,t,\zeta)$ of the target
proton \cite{Ji:1996kb,Ji:1997ek,Radyushkin:1996nd,Ji:1998xh,%
Guichon:1998xv,Vanderhaeghen:1998uc,Radyushkin:1999es,%
Collins:1999be,Diehl:1999tr,Diehl:1999kh,Blumlein:2000cx,Penttinen:2000dg}.
Here $x$ is the light-front momentum fraction of the struck quark,
and $\zeta= Q^2/2 P\cdot q$ plays the role of the Bjorken
variable.  The square of the four-momentum transfer from the
proton is given by $t=\Delta^2\ = \ 2P\cdot \Delta\  =\
-{(\zeta^2M^2+{\vec \Delta_\perp}^2)\over (1-\zeta)}\ $ , where
$\Delta $ is the difference of initial and final momenta of the
proton ($P=P'+\Delta$). We will be interested in deeply virtual
Compton scattering where $q^2$ is large compared to the masses and
$t$.  Then, to leading order in $1/Q^2$, ${-q^2\over 2P_I\cdot
q}=\zeta\ .$ Thus $\zeta$ plays the role of the Bjorken variable
in deeply virtual Compton scattering. For a fixed value of $-t$,
the allowed range of $\zeta$ is given by
\begin{equation}
0\ \le\ \zeta\ \le\ {(-t)\over 2M^2}\ \ \left( {\sqrt{1+{4M^2\over
(-t)}}}\ -\ 1\ \right)\ . \label{nn4}
\end{equation}
The form factor $H(x,t,\zeta)$ describes the proton response when
the helicity of the proton is unchanged, and $E(x,t,\zeta)$ is for
the case when the proton helicity is flipped.  Two additional
functions $\tilde H(x,t,\zeta),$ and $\tilde E(x,t,\zeta)$ appear,
corresponding to the dependence of the Compton amplitude on quark
helicity.

\begin{figure}
\begin{center}
\includegraphics{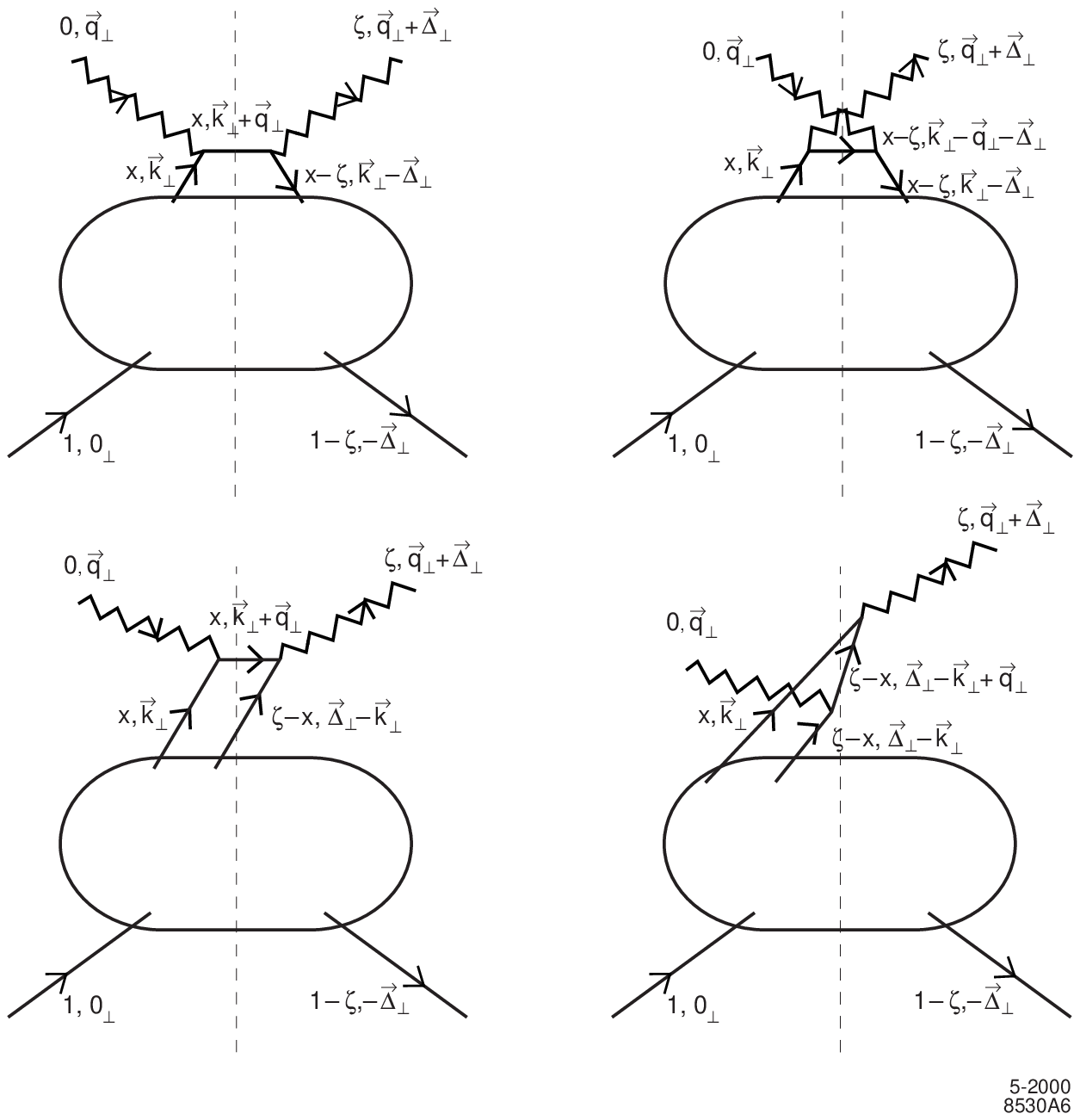}
\end{center}
\caption[*]{Light-front time-ordered contributions to deeply
virtual Compton scattering.  Only the contributions of leading
twist in $1/q^2$ are illustrated.  These contributions illustrate
the factorization property of the leading twist amplitude.
\label{fig:3}}
\end{figure}

Recently, Markus Diehl, Dae Sung Hwang and I \cite{Brodsky:2000xy}
have shown how the deeply virtual Compton amplitude can be
evaluated explicitly in the Fock state representation using the
matrix elements of the currents and the boost properties of the
light-front wavefunctions. For the $n \to n$ diagonal term
($\Delta n = 0$), the arguments of the final-state hadron
wavefunction are $x_1-\zeta \over 1-\zeta$, ${\vec{k}}_{\perp 1} -
{1-x_1\over 1-\zeta} {\vec{\Delta}}_\perp$ for the struck quark
and $x_i\over 1-\zeta$, ${\vec{k}}_{\perp i} + {x_i\over 1-\zeta}
{\vec{\Delta}}_\perp$ for the $n-1$ spectators. We thus obtain
formulae for the diagonal (parton-number-conserving) contribution
to the generalized form factors for deeply virtual Compton
amplitude in the domain
\cite{Diehl:1999kh,Diehl:1999tr,Muller:1994fv}
$\zeta\le x_1\le 1$:\\
\begin{eqnarray}
&&{\sqrt{1-\zeta}}f_{1\, (n\to n)}(x_1,t,\zeta)\, -\,
{\zeta^2\over 4{\sqrt{1-\zeta}}} f_{2\, (n\to n)}(x_1,t,\zeta)
\nonumber\\
 &=&
\sum_{n, ~ \lambda} \prod_{i=1}^{n} \int^1_0 dx_{i(i\ne 1)} \int
{d^2{\vec{k}}_{\perp i} \over 2 (2\pi)^3 } ~
\delta\left(1-\sum_{j=1}^n x_j\right) ~ \delta^{(2)}
\left(\sum_{j=1}^n {\vec{k}}_{\perp j}\right)  \nonumber\\[1ex]
&&\times \psi^{\uparrow \ *}_{(n)}(x^\prime_i,
{\vec{k}}^\prime_{\perp i},\lambda_i) ~ \psi^{\uparrow}_{(n)}(x_i,
{\vec{k}}_{\perp i},\lambda_i) (\sqrt{1-\zeta})^{1-n}, \label{t1}
\end{eqnarray}
\begin{eqnarray}
&& {\sqrt{1-\zeta}}\,\left(\, 1+{\zeta\over 2(1-\zeta)}\,\right)\,
{(\Delta^1-{\mathrm i} \Delta^2)\over 2M}f_{2\, (n\to
n)}(x_1,t,\zeta)
\nonumber\\
 &=&
\sum_{n, ~ \lambda} \prod_{i=1}^{n} \int^1_0 dx_{i(i\ne 1)} \int
{d^2{\vec{k}}_{\perp i} \over 2 (2\pi)^3 } ~
\delta\left(1-\sum_{j=1}^n x_j\right) ~ \delta^{(2)}
\left(\sum_{j=1}^n {\vec{k}}_{\perp j}\right)  \nonumber\\[1ex]
&&\qquad\qquad\qquad\times \psi^{\uparrow \ *}_{(n)}(x^\prime_i,
{\vec{k}}^\prime_{\perp i},\lambda_i) ~
\psi^{\downarrow}_{(n)}(x_i, {\vec{k}}_{\perp i},\lambda_i)
(\sqrt{1-\zeta})^{1-n} , \label{t1f2}
\end{eqnarray}
where
\begin{equation}
\left\{ \begin{array}{lll} x^\prime_1 = {x_1-\zeta \over
1-\zeta}\, ,\ &{\vec{k}}^\prime_{\perp 1} ={\vec{k}}_{\perp 1} -
{1-x_1\over 1-\zeta} {\vec{\Delta}}_\perp
&\mbox{for the struck quark,}\\[1ex]
x^\prime_i = {x_i\over 1-\zeta}\, ,\ &{\vec{k}}^\prime_{\perp i}
={\vec{k}}_{\perp i} + {x_i\over 1-\zeta} {\vec{\Delta}}_\perp
&\mbox{for the $ (n-1)$ spectators.}
\end{array}\right.
\label{t2}
\end{equation}
A sum over all possible helicities $\lambda_i$ is understood. If
quark masses are neglected, the currents conserve helicity. We
also can check that $\sum_{i=1}^n x^\prime_i = 1$, $\sum_{i=1}^n
{\vec{k}}^\prime_{\perp i} = {\vec{0}}_\perp$.

For the $n+1 \to n-1$ off-diagonal term ($\Delta n = -2$),
consider the case where partons $1$ and $n+1$ of the initial
wavefunction annihilate into the current leaving $n-1$ spectators.
Then $x_{n+1} = \zeta - x_{1}$, ${\vec{k}}_{\perp n+1} =
{\vec{\Delta}}_\perp-{\vec{k}}_{\perp 1}$. The remaining $n-1$
partons have total momentum $((1-\zeta)P^+,
-{\vec{\Delta}}_{\perp})$. The final wavefunction then has
arguments $x^\prime_i = {x_i \over 1- \zeta}$ and
${\vec{k}}^\prime_{\perp i}= {\vec{k}}_{\perp i} + {x_i\over
1-\zeta} {\vec{\Delta}}_\perp$. We thus obtain the formulae for
the off-diagonal matrix element
of the Compton amplitude in the domain $0\le x_1\le \zeta$:\\
\begin{eqnarray}
&&{\sqrt{1-\zeta}}f_{1\, (n+1\to n-1)}(x_1,t,\zeta)\, -\,
{\zeta^2\over 4{\sqrt{1-\zeta}}} f_{2\, (n+1\to n-1)}(x_1,t,\zeta)
\nonumber\\
 &=&
\sum_{n, ~ \lambda} \int^1_0 dx_{n+1} \int {d^2{\vec{k}}_{\perp 1}
\over 2 (2\pi)^3 } \int {d^2{\vec{k}}_{\perp n+1} \over 2 (2\pi)^3
} \prod_{i=2}^{n} \int^1_0 dx_{i} \int {d^2{\vec{k}}_{\perp i}
\over 2 (2\pi)^3 }
\nonumber\\[2ex]
&&\times \delta\left(1-\sum_{j=1}^{n+1} x_j\right) ~
\delta^{(2)}\left(\sum_{j=1}^{n+1} {\vec{k}}_{\perp j}\right)
[\sqrt{1-\zeta}]^{1-n}
\nonumber\\[2ex]
&&\times\psi^{\uparrow\ *}_{(n-1)}
(x^\prime_i,{\vec{k}}^\prime_{\perp i},\lambda_i) ~
\psi^{\uparrow}_{(n+1)}(\{x_1, x_i, x_{n+1} = \zeta - x_{1}\},
\nonumber\\[2ex]
&&\qquad\ \ \{ {\vec{k}}_{\perp 1}, {\vec{k}}_{\perp i},
{\vec{k}}_{\perp n+1} = {\vec{\Delta}}_\perp-{\vec{k}}_{\perp
1}\}, \{\lambda_1,\lambda_{i},\lambda_{n+1} = - \lambda_{1}\}) ,
\end{eqnarray}
\begin{eqnarray}
&& {\sqrt{1-\zeta}}\,\Big(\, 1+{\zeta\over 2(1-\zeta)}\,\Big)\,
{(\Delta^1-{\mathrm i} \Delta^2)\over 2M}f_{2\, (n+1\to
n-1)}(x_1,t,\zeta)
\nonumber\\
 &=&
\sum_{n, ~ \lambda} \int^1_0 dx_{n+1} \int {d^2{\vec{k}}_{\perp 1}
\over 2 (2\pi)^3 } \int {d^2{\vec{k}}_{\perp n+1} \over 2 (2\pi)^3
} \prod_{i=2}^{n} \int^1_0 dx_{i} \int {d^2{\vec{k}}_{\perp i}
\over 2 (2\pi)^3 }
\nonumber\\[2ex]
&&\qquad\qquad\qquad\times \delta\left(1-\sum_{j=1}^{n+1}
x_j\right) ~ \delta^{(2)}\left(\sum_{j=1}^{n+1} {\vec{k}}_{\perp
j}\right) [\sqrt{1-\zeta}]^{1-n}
\nonumber\\[2ex]
&&\qquad\qquad\qquad\times\psi^{\uparrow\ *}_{(n-1)}
(x^\prime_i,{\vec{k}}^\prime_{\perp i},\lambda_i) ~
\psi^{\downarrow}_{(n+1)}(\{x_1, x_i, x_{n+1} = \zeta - x_{1}\},
\nonumber\\[2ex]
&&\qquad\qquad\ \ \{ {\vec{k}}_{\perp 1}, {\vec{k}}_{\perp i},
{\vec{k}}_{\perp n+1} = {\vec{\Delta}}_\perp-{\vec{k}}_{\perp
1}\}, \{\lambda_1,\lambda_{i},\lambda_{n+1} = - \lambda_{1}\}) ,
\end{eqnarray}
where $i=2,3,\cdots ,n$ label the $n-1$ spectator partons which
appear in the final-state hadron wavefunction with
\begin{equation}
x^\prime_i = {x_i\over 1-\zeta}\, ,\qquad {\vec{k}}^\prime_{\perp
i} ={\vec{k}}_{\perp i} + {x_i\over 1-\zeta} {\vec{\Delta}}_\perp
\ .
\end{equation}
We can again check that the arguments of the final-state
wavefunction satisfy $\sum_{i=2}^n x^\prime_i = 1$, $\sum_{i=2}^n
{\vec{k}}^\prime_{\perp i} = {\vec{0}}_\perp$.

The above representation is the general form for the generalized
form factors of the deeply virtual Compton amplitude for any
composite system. Thus given the light-front Fock state
wavefunctions of the eigensolutions of the light-front
Hamiltonian, we can compute the amplitude for virtual Compton
scattering including all spin correlations.  The formulae are
accurate to leading order in $1/Q^2$.  Radiative corrections to
the quark Compton amplitude of order $\alpha_s(Q^2)$ from diagrams
in which a hard gluon interacts between the two photons have also
been neglected.

\section{Applications of QCD Factorization to Hard QCD
 Processes}

Factorization theorems for hard exclusive, semi-exclusive, and
diffractive processes allow the separation of soft
non-perturbative dynamics of the bound state hadrons from the hard
dynamics of a perturbatively-calculable quark-gluon scattering
amplitude.  The factorization of inclusive reactions is reviewed
in ref. For reviews and bibliography of exclusive process
calculations in QCD (see Ref.
\cite{Brodsky:1989pv,Brodsky:2000dr}).

The light-front formalism provides a physical factorization scheme
which conveniently separates and factorizes soft non-perturbative
physics from hard perturbative dynamics in both exclusive and
inclusive reactions \cite{Lepage:1980fj,Lepage:1979zb}.

In hard inclusive reactions all intermediate states are divided
according to $\M^2_n < \Lambda^2 $ and $\M^2_n > \Lambda^2 $
domains.  The lower mass regime is associated with the quark and
gluon distributions defined from the absolute squares of the LC
wavefunctions in the light front factorization scheme.  In the
high invariant mass regime, intrinsic transverse momenta can be
ignored, so that the structure of the process at leading power has
the form of hard scattering on collinear quark and gluon
constituents, as in the parton model.  The attachment of gluons
from the LC wavefunction to a propagator in a hard subprocess is
power-law suppressed in LC gauge, so that the minimal quark-gluon
particle-number subprocesses dominate.  It is then straightforward
to derive the DGLAP equations from the evolution of the
distributions with $\log \Lambda^2$. The anomaly contribution to
singlet helicity structure function $g_1(x,Q)$ can be explicitly
identified in the LC factorization scheme as due to the $\gamma^*
g \to q \bar q$ fusion process.  The anomaly contribution would be
zero if the gluon is on shell.  However, if the off-shellness of
the state is larger than the quark pair mass, one obtains the
usual anomaly contribution \cite{Bass:1999rn}.

In exclusive amplitudes, the LC wavefunctions are the
interpolating amplitudes connecting the quark and gluons to the
hadronic states.  In an exclusive amplitude involving a hard scale
$Q^2$ all intermediate states can be divided according to $\M^2_n
< \Lambda^2 < Q^2 $ and $\M^2_n < \Lambda^2 $ invariant mass
domains.  The high invariant mass contributions to the amplitude
has the structure of a hard scattering process $T_H$ in which the
hadrons are replaced by their respective (collinear) quarks and
gluons.  In light-cone gauge only the minimal Fock states
contribute to the leading power-law fall-off of the exclusive
amplitude. The wavefunctions in the lower invariant mass domain
can be integrated up to an arbitrary intermediate invariant mass
cutoff $\Lambda$.  The invariant mass domain beyond this cutoff is
included in the hard scattering amplitude $T_H$.  The $T_H$
satisfy dimensional counting rules \cite{Brodsky:1975vy}.
Final-state and initial state corrections from gluon attachments
to lines connected to the color-singlet distribution amplitudes
cancel at leading twist.  Explicit examples of perturbative QCD
factorization will be discussed in more detail in the next
section.

The key non-perturbative input for exclusive processes is thus the
gauge and frame independent hadron distribution amplitude
\cite{Lepage:1979zb,Lepage:1980fj} defined as the integral of the
valence (lowest particle number) Fock wavefunction; \eg\ for the
pion
\begin{equation}
\phi_\pi (x_i,\Lambda) \equiv \int d^2k_\perp\,
\psi^{(\Lambda)}_{q\bar q/\pi} (x_i, \vec k_{\perp i},\lambda)
\label{eq:f1a}
\end{equation}
where the global cutoff $\Lambda$ is identified with the
resolution $Q$. The distribution amplitude controls leading-twist
exclusive amplitudes at high momentum transfer, and it can be
related to the gauge-invariant Bethe-Salpeter wavefunction at
equal light-front time.  The logarithmic evolution of hadron
distribution amplitudes $\phi_H (x_i,Q)$ can be derived from the
perturbatively-computable tail of the valence light-front
wavefunction in the high transverse momentum regime
\cite{Lepage:1979zb,Lepage:1980fj}. The conformal basis for the
evolution of the three-quark distribution amplitudes for the
baryons~\cite{Lepage:1979za} has recently been obtained by V.
Braun \etal \cite{Braun:1999te}.

The existence of an exact formalism provides a basis for
systematic approximations and a control over neglected terms.  For
example, one can analyze exclusive semi-leptonic $B$-decays which
involve hard internal momentum transfer using a perturbative QCD
formalism \cite{Szczepaniak:1990dt,Szczepaniak:1996xg,%
Beneke:1999br,Keum:2000ph,Keum:2000wi,Li:2000hh} patterned after
the perturbative analysis of form factors at large momentum
transfer.  The hard-scattering analysis again proceeds by writing
each hadronic wavefunction as a sum of soft and hard contributions
\begin{equation}
\psi_n = \psi^{{\rm soft}}_n (\M^2_n < \Lambda^2) + \psi^{{\rm
hard}}_n (\M^2_n >\Lambda^2) ,
\end{equation}
where $\M^2_n $ is the invariant mass of the partons in the
$n$-particle Fock state and $\Lambda$ is the separation scale. The
high internal momentum contributions to the wavefunction
$\psi^{{\rm hard}}_n $ can be calculated systematically from QCD
perturbation theory by iterating the gluon exchange kernel.  The
contributions from high momentum transfer exchange to the
$B$-decay amplitude can then be written as a convolution of a
hard-scattering quark-gluon scattering amplitude $T_H$ with the
distribution amplitudes $\phi(x_i,\Lambda)$, the valence
wavefunctions obtained by integrating the constituent momenta up
to the separation scale ${\cal M}_n < \Lambda < Q$.  Furthermore
in processes such as $B \to \pi D$ where the pion is effectively
produced as a rapidly-moving small Fock state with a small
color-dipole interactions,  final state interactions are
suppressed by color transparency.  This is the basis for the
perturbative hard-scattering
analyses \cite{Szczepaniak:1990dt,Beneke:1999br,Keum:2000ph,%
Keum:2000wi,Li:2000hh}. In a systematic analysis, one can identify
the hard PQCD contribution as well as the soft contribution from
the convolution of the light-front wavefunctions. Furthermore, the
hard-scattering contribution can be systematically improved.

Given the solution for the hadronic wavefunctions
$\psi^{(\Lambda)}_n$ with $\M^2_n < \Lambda^2$, one can construct
the wavefunction in the hard regime with $\M^2_n > \Lambda^2$
using projection operator techniques.  The construction can be
done perturbatively in QCD since only high invariant mass, far
off-shell matrix elements are involved.  One can use this method
to derive the physical properties of the LC wavefunctions and
their matrix elements at high invariant mass.  Since $\M^2_n =
\sum^n_{i=1} \left(\frac{k^2_\perp+m^2}{x}\right)_i $, this method
also allows the derivation of the asymptotic behavior of
light-front wavefunctions at large $k_\perp$, which in turn leads
to predictions for the fall-off of form factors and other
exclusive matrix elements at large momentum transfer, such as the
quark counting rules for predicting the nominal power-law fall-off
of two-body scattering amplitudes at fixed $\theta_{cm}$
\cite{Brodsky:1975vy} and helicity selection rules
\cite{Brodsky:1981kj}. The phenomenological successes of these
rules can be understood within QCD if the coupling $\alpha_V(Q)$
freezes in a range of relatively small momentum transfer
\cite{Brodsky:1998dh}.

\section{Two-Photon Processes}

The simplest and perhaps the most elegant illustration of an
exclusive reaction in QCD is the evaluation of the photon-to-pion
transition form factor $F_{\gamma \to \pi}(Q^2)$
\cite{Lepage:1980fj,Brodsky:1981rp} which is measurable in
single-tagged two-photon $ee \to ee \pi^0$ reactions. The form
factor is defined via the invariant amplitude $ \Gamma^\mu = -ie^2
F_{\pi \gamma}(Q^2) \epsilon^{\mu \nu \rho \sigma} p^\pi_\nu
\epsilon_\rho q_\sigma \ .$ As in inclusive reactions, one must
specify a factorization scheme which divides the integration
regions of the loop integrals into hard and soft momenta, compared
to the resolution scale $\tilde Q$. At leading twist, the
transition form factor then factorizes as a convolution of the
$\gamma^* \gamma \to q \bar q$ amplitude (where the quarks are
collinear with the final state pion) with the valence light-front
wavefunction of the pion:
\begin{equation}
F_{\gamma M}(Q^2)= {4 \over \sqrt 3}\int^1_0 dx \phi_M(x,\tilde Q)
T^H_{\gamma \to M}(x,Q^2) . \label{transitionformfactor}
\end{equation}
The hard scattering amplitude for $\gamma\gamma^*\to q \bar q$ is
$ T^H_{\gamma M}(x,Q^2) = { [(1-x) Q^2]^{-1}}\times\break \left(1
+ {\cal O}(\alpha_s)\right). $ The leading QCD corrections have
been computed by Braaten \cite{Braaten:1987yy}. The evaluation of
the next-to-leading corrections in the physical $\alpha_V$ scheme
is given in Ref. \cite{Brodsky:1998dh}. For the asymptotic
distribution amplitude $\phi^{\rm asympt}_\pi (x) = \sqrt 3 f_\pi
x(1-x)$ one predicts $ Q^2 F_{\gamma \pi}(Q^2)= 2 f_\pi \left(1 -
{5\over3} {\alpha_V(Q^*)\over \pi}\right)$ where $Q^*= e^{-3/2} Q$
is the BLM scale for the pion form factor.  The PQCD predictions
have been tested in measurements of $e \gamma \to e \pi^0$ by the
CLEO collaboration \cite{Gronberg:1998fj}. See Fig.
\ref{Fig:DalleyCleo} (b). The observed flat scaling of the $Q^2
F_{\gamma \pi}(Q^2)$ data from $Q^2 = 2$ to $Q^2 = 8$ GeV$^2$
provides an important confirmation of the applicability of leading
twist QCD to this process.  The magnitude of $Q^2 F_{\gamma
\pi}(Q^2)$ is remarkably consistent with the predicted form,
assuming the asymptotic distribution amplitude and including the
LO QCD radiative correction with $\alpha_V(e^{-3/2} Q)/\pi \simeq
0.12$.  One could allow for some broadening of the distribution
amplitude with a corresponding increase in the value of $\alpha_V$
at small scales. Radyushkin \cite{Radyushkin:1995pj}, Ong
\cite{Ong:1995gs}, and Kroll \cite{Kroll:1996jx} have also noted
that the scaling and normalization of the photon-to-pion
transition form factor tends to favor the asymptotic form for the
pion distribution amplitude and rules out broader distributions
such as the two-humped form suggested by QCD sum rules
\cite{Chernyak:1984ej}.

\begin{figure}
\begin{center}
\includegraphics[width=5in]{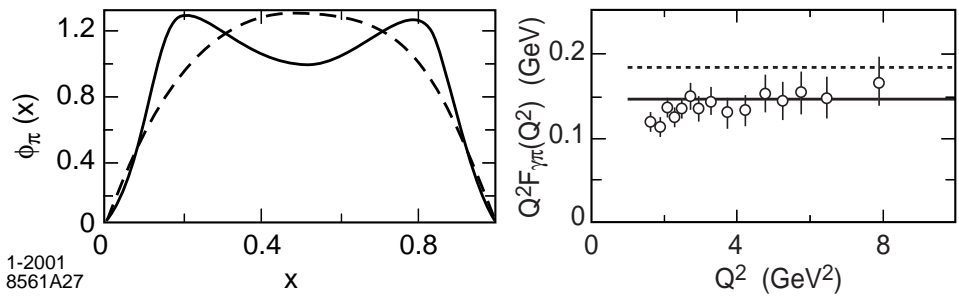}
\end{center}
\caption[*]{ (a) Preliminary transverse lattice results for the
pion distribution amplitude at $Q^2 \sim 10 ~{\rm GeV}^2$.  The
solid curve is the theoretical prediction from the combined
DLCQ/transverse lattice method \cite{Dalley:2000dh}; the chain
line is the experimental result obtained from jet diffractive
dissociation \cite{Ashery:1999nq}.  Both are normalized to the
same area for comparison. (b) Scaling of the transition photon to
pion transition form factor $Q^2F_{\gamma \pi^0}(Q^2)$.  The
dotted and solid theoretical curves are the perturbative QCD
prediction at leading and next-to-leading order, respectively,
assuming the asymptotic pion distribution The data are from the
CLEO collaboration \cite{Gronberg:1998fj}. \label{Fig:DalleyCleo}}
\end{figure}

The two-photon annihilation process $\gamma^* \gamma \to $
hadrons, which is measurable in single-tagged $e^+ e^- \to e^+ e^-
{\rm hadrons}$ events, provides a semi-local probe of $C=+$ hadron
systems $\pi^0, \eta^0, \eta^\prime, \eta_c, \pi^+ \pi^-$, etc.
The $\gamma^* \gamma \to \pi^+ \pi^-$ hadron pair process is
related to virtual Compton scattering on a pion target by
crossing.  The leading twist amplitude is sensitive to the $1/x -
1/(1-x)$ moment of the two-pion distribution amplitude coupled to
two valence quarks \cite{Muller:1994fv,Diehl:2000uv}.

Two-photon reactions, $\gamma \gamma \to H \bar H$ at large s =
$(k_1 + k_2)^2$ and fixed $\theta_{\rm cm}$, provide a
particularly important laboratory for testing QCD since these
cross-channel ``Compton" processes are the simplest calculable
large-angle exclusive hadronic scattering reactions. The helicity
structure, and often even the absolute normalization can be
rigorously computed for each two-photon channel
\cite{Brodsky:1981rp}. In the case of meson pairs, dimensional
counting predicts that for large $s$, $s^4 d\sigma/dt(\gamma
\gamma \to M \bar M$ scales at fixed $t/s$ or $\theta_{\rm c.m.}$
up to factors of $\ln s/\Lambda^2$. The angular dependence of the
$\gamma \gamma \to H \bar H$ amplitudes can be used to determine
the shape of the process-independent distribution amplitudes,
$\phi_H(x,Q)$. An important feature of the $\gamma \gamma \to M
\bar M$ amplitude for meson pairs is that the contributions of
Landshoff pitch singularities are power-law suppressed at the Born
level---even before taking into account Sudakov form factor
suppression.  There are also no anomalous contributions from the
$x \to 1$ endpoint integration region. Thus, as in the calculation
of the meson form factors, each fixed-angle helicity amplitude can
be written to leading order in $1/Q$ in the factorized form $[Q^2
= p_T^2 = tu/s; \tilde Q_x = \min(xQ,(l-x)Q)]$:
\begin{equation}{\cal M}_{\gamma \gamma\to M \bar M}
= \int^1_0 dx \int^1_0 dy \phi_{\bar M}(y,\tilde Q_y)
T_H(x,y,s,\theta_{\rm c.m.} \phi_{M}(x,\tilde Q_x) ,
\end{equation}
where $T_H$ is the hard-scattering amplitude $\gamma \gamma \to (q
\bar q) (q \bar q)$ for the production of the valence quarks
collinear with each meson, and $\phi_M(x,\tilde Q)$ is the
amplitude for finding the valence $q$ and $\bar q$ with
light-front fractions of the meson's momentum, integrated over
transverse momenta $k_\perp < \tilde Q.$ The contribution of
non-valence Fock states are power-law suppressed. Furthermore, the
helicity-selection rules \cite{Brodsky:1981kj} of perturbative QCD
predict that vector mesons are produced with opposite helicities
to leading order in $1/Q$ and all orders in $\alpha_s$. The
dependence in $x$ and $y$ of several terms in $T_{\lambda,
\lambda'}$ is quite similar to that appearing in the meson's
electromagnetic form factor. Thus much of the dependence on
$\phi_M(x,Q)$ can be eliminated by expressing it in terms of the
meson form factor. In fact, the ratio of the $\gamma \gamma \to
\pi^+ \pi^-$ and $e^+ e^- \to \mu^+ \mu^-$ amplitudes at large $s$
and fixed $\theta_{CM}$ is nearly insensitive to the running
coupling and the shape of the pion distribution amplitude:
\begin{equation}
{{d\sigma \over dt }(\gamma \gamma \to \pi^+ \pi^-) \over {d\sigma
\over dt }(\gamma \gamma \to \mu^+ \mu^-)} \sim {4 \vert F_\pi(s)
\vert^2 \over 1 - \cos^2 \theta_{\rm c.m.} } .
\end{equation}
The comparison of the PQCD prediction for the sum of $\pi^+ \pi^-$
plus $K^+ K^-$ channels with recent CLEO data \cite{Paar} is shown
in Fig. \ref{Fig:CLEO}. The CLEO data for charged pion and kaon
pairs show a clear transition to the scaling and angular
distribution predicted by PQCD \cite{Brodsky:1981rp} for $W =
\sqrt(s_{\gamma \gamma} > 2$ GeV.  See Fig. \ref{Fig:CLEO}.  It is
clearly important to measure the magnitude and angular dependence
of the two-photon production of neutral pions and $\rho^+ \rho^-$
cross sections in view of the strong sensitivity of these channels
to the shape of meson distribution amplitudes. QCD also predicts
that the production cross section for charged $\rho$-pairs (with
any helicity) is much larger that for that of neutral $\rho$
pairs, particularly at large $\theta_{\rm c.m.}$ angles. Similar
predictions are possible for other helicity-zero mesons. The cross
sections for Compton scattering on protons and the crossed
reaction $\gamma \gamma \to p \bar p$ at high momentum transfer
have also been evaluated \cite{Farrar:1990qj,Brooks:2000nb},
providing important tests of the proton distribution amplitude.

\begin{figure}
\begin{center}
\includegraphics[width=5in]{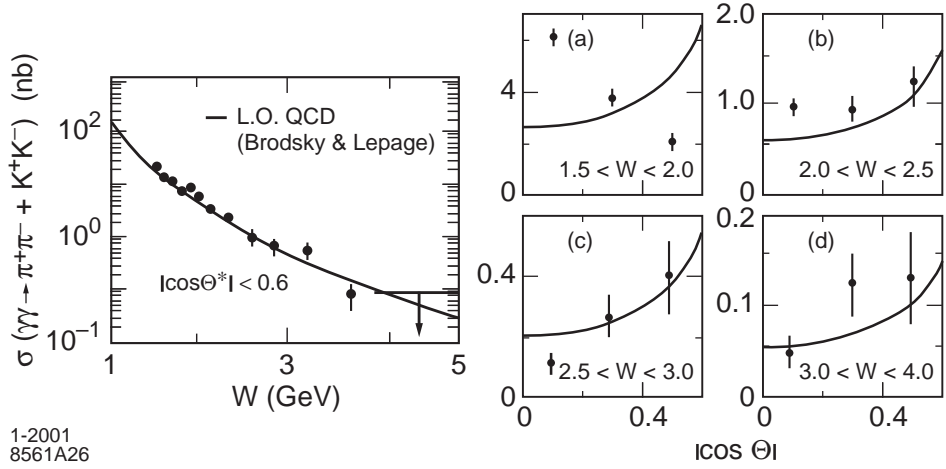}
\end{center}
\caption[*]{Comparison of the sum of $\gamma \gamma \rightarrow
\pi^+ \pi^-$ and $\gamma \gamma \rightarrow K^+ K^-$ meson pair
production cross sections with the scaling and angular
distribution of the perturbative QCD prediction
\cite{Brodsky:1981rp}.  The data are from the CLEO collaboration
\cite{Paar}. \label{Fig:CLEO}}
\end{figure}

It is particularly compelling to see a transition in angular
dependence between the low energy chiral and PQCD regimes.  The
success of leading-twist perturbative QCD scaling for exclusive
processes at presently experimentally accessible momentum transfer
can be understood if the effective coupling $\alpha_V(Q^*)$ is
approximately constant at the relatively small scales $Q^*$
relevant to the hard scattering amplitudes \cite{Brodsky:1998dh}.
The evolution of the quark distribution amplitudes In the
low-$Q^*$ domain at also needs to be minimal.  Sudakov suppression
of the endpoint contributions is also strengthened if the coupling
is frozen because of the exponentiation of a double logarithmic
series.

A debate has continued
\cite{Isgur:1989cy,Radyushkin:1998rt,Bolz:1996sw,Vogt:2000bz} on
whether processes such as the pion and proton form factors and
elastic Compton scattering $\gamma p \to \gamma p$ might be
dominated by higher-twist mechanisms until very large momentum
transfer.  If one assumes that the light-front wavefunction of the
pion has the form $\psi_{\rm soft}(x,k_\perp) = A \exp (-b
{k_\perp^2\over x(1-x)})$, then the Feynman endpoint contribution
to the overlap integral at small $k_\perp$ and $x \simeq 1$ will
dominate the form factor compared to the hard-scattering
contribution until very large $Q^2$.  However, this ansatz for
$\psi_{\rm soft}(x,k_\perp)$ has no suppression at $k_\perp =0$
for any $x$; \ie, the wavefunction in the hadron rest frame does
not fall-off at all for $k_\perp = 0$ and $k_z \to - \infty$. Thus
such wavefunctions do not represent well soft QCD contributions.
Endpoint contributions are also suppressed by the QCD Sudakov form
factor, reflecting the fact that a near-on-shell quark must
radiate if it absorbs large momentum. One can show
\cite{Lepage:1980fj} that the leading power dependence of the
two-particle light-front Fock wavefunction in the endpoint region
is $1-x$, giving a meson structure function which falls as
$(1-x)^2$ and thus by duality a non-leading contribution to the
meson form factor $F(Q^2) \propto 1/Q^3$.  Thus the dominant
contribution to meson form factors comes from the hard-scattering
regime.

Radyushkin \cite{Radyushkin:1998rt} has argued that the Compton
amplitude is dominated by soft end-point contributions of the
proton wavefunctions where the two photons both interact on a
quark line carrying nearly all of the proton's momentum.  This
description appears to agree with the Compton data at least at
forward angles where $-t < 10$ GeV$^2$.  From this viewpoint, the
dominance of the factorizable PQCD leading twist contributions
requires momentum transfers much higher than those currently
available.  However, the endpoint model cannot explain the
empirical success of the perturbative QCD fixed $\theta_{c.m.}$
scaling $s^7 d\sigma/dt(\gamma p \to \pi^+ n) \sim {\rm const} $
at relatively low momentum transfer in pion photoproduction
\cite{Anderson:1973cc}.

Clearly much more experimental input on hadron wavefunctions is
needed, particularly from measurements of two-photon exclusive
reactions into meson and baryon pairs at the high luminosity $B$
factories.  For example, the ratio \hfill\break ${{d\sigma \over
dt }(\gamma \gamma \to \pi^0 \pi^0) / {d\sigma \over dt}(\gamma
\gamma \to \pi^+ \pi^-)}$ is particularly sensitive to the shape
of pion distribution amplitude. Baryon pair production in
two-photon reactions at threshold may reveal physics associated
with the soliton structure of baryons in QCD
\cite{Sommermann:1992yh,Marek:2001}.  In addition, fixed target
experiments can provide much more information on fundamental QCD
processes such as deeply virtual Compton scattering and large
angle Compton scattering.

\section{Diffractive Dissociation and Light-Cone Wavefunctions}

Diffractive dissociation in QCD can be understood as a three-step
process:

1.  The initial hadron can be decomposed in terms of its quark and
gluon constituents in terms of its light-front Fock-state
components.

2.  In the second step, the incoming hadron is resolved by Pomeron
or Odderon (multi-gluon) exchange with the target or by Coulomb
dissociation. The exchanged interaction has to supply sufficient
momentum transfer $q^\mu$ to put the diffracted state $X$ on
shell.  Light-front energy conservation requires $q^- = {( m_X^2-
m_\pi^2)/ P^+_\pi},$ where $m_X$ is the invariant mass of $X$.  In
a heavy target rest system, the longitudinal momentum transfer is
$q^z = {(m_X^2- m_\pi^2)/ E_{\pi {\rm lab}}}.$ Thus the momentum
transfer $t = q^2$ to the target can be sufficiently small so that
the target remains intact.

In perturbative QCD, the pomeron is generally be represented as
multiple gluon exchange between the target and projectile.
Effectively this interaction occurs over a short light-front time
interval, and thus like photon exchange, the perturbative QCD
pomeron can be effectively represented as a local operator.  This
description is believed to be applicable when the pomeron has to
resolve compact states and is the basis for the terminology ``hard
pomeron".  The BFKL formalism generalizes the perturbative QCD
treatment by an all-orders perturbative resummation, generating a
pomeron with a fixed Regge intercept $\alpha_P(0)$.  Next to
leading order calculations with BLM scale fixing leads to a
predicted intercept $\alpha_P(0) \simeq
0.4$~\cite{Brodsky:1999kn}.   However, when the exchange
interactions are soft, a multiperipheral description in terms of
meson ladders may dominate the physics.  This is the basis for the
two-component pomeron model of Donnachie and
Landshoff~\cite{Donnachie:2001xx}.

Consider a collinear frame where the incident momentum $P^+_\pi$
is large and $s = (p_\pi + p_{\rm target})^2 \simeq p^+_\pi
p^-_{\rm target}.$ The matrix element of an exchanged gluon with
momentum $q_i$ between the projectile and an intermediate state
$\ket N$ is dominated by the ``plus current": $\VEV{\pi|j^+(0)\exp
(i {\half q^+_i x^-}-i q_{\perp i} \cdot x_\perp|N}$. Note that
the coherent sum of couplings of an exchanged gluon to the pion
system vanishes when its momentum is small compared to the
characteristic momentum scales in the projectile light-front
wavefunction: $q^{\perp_i} \Delta x_\perp \ll 1$ and $q^+_i \Delta
x^- \ll 1$.  The destructive interference of the gauge couplings
to the constituents of the projectile follows simply from the fact
that the color charge operator has zero matrix element between
distinct eigenstates of the QCD Hamiltonian: $\VEV{A|Q|B} \equiv
\int d^2x{_\perp} dx^- \VEV{A|j^+(0)|B} = 0$~\cite{BH}. At high
energies the change in $k^+_i$ of the constituents can be ignored,
so that Fock states of a hadron with small transverse size
interact weakly even in a nuclear target because of their small
dipole moment~\cite{Brodsky:1988xz,Bertsch:1981py}.   To a good
approximation the sum of couplings to the constituents of the
projectile can be represented as a derivative with respect to
transverse momentum. Thus photon exchange measures a weighted sum
of transverse derivatives $\partial_{k_\perp} \psi_n(x_i,
k_{\perp_i},\lambda_i),$ and two-gluon exchange measures the
second transverse partial derivative~\cite{BHDP}.

3.  The final step is the hadronization of the $n$ constituents of
the projectile Fock state into final state hadrons.  Since $q^+_i$
is small, the number of partons in the initial Fock state and the
final state hadrons are unchanged.  Their coalescence is thus
governed by the convolution of initial and final-state Fock state
wavefunctions.  In the case of states with high $k_\perp$, the
final state will hadronize into jets, each reflecting the
respective $x_i$ of the Fock state constituents. In the case of
higher Fock states with intrinsic sea quarks such as an extra $c
\bar c$ pair (intrinsic charm), one will observe leading $J/\psi$
or open charm hadrons in the projectile fragmentation region; \ie,
the hadron's fragments will tend to have the same rapidity as that
of the projectile.

For example, diffractive multi-jet production in heavy nuclei
provides a novel way to measure the shape of the LC Fock state
wavefunctions and test color transparency.  Consider the reaction
\cite{Bertsch:1981py,Frankfurt:1993it,Frankfurt:2000jm} $\pi A
\rightarrow {\rm Jet}_1 + {\rm Jet}_2 + A^\prime$ at high energy
where the nucleus $A^\prime$ is left intact in its ground state.
The transverse momenta of the jets balance so that $ \vec k_{\perp
i} + \vec k_{\perp 2} = \vec q_\perp < {R^{-1}}_A \ . $ The
light-front longitudinal momentum fractions also need to add to
$x_1+x_2 \sim 1$ so that $\Delta p_L < R^{-1}_A$.  The process can
then occur coherently in the nucleus. Because of color
transparency, the valence wavefunction of the pion with small
impact separation, will penetrate the nucleus with minimal
interactions, diffracting into jet pairs \cite{Bertsch:1981py}.
The $x_1=x$, $x_2=1-x$ dependence of the di-jet distributions will
thus reflect the shape of the pion valence light-front
wavefunction in $x$; similarly, the $\vec k_{\perp 1}- \vec
k_{\perp 2}$ relative transverse momenta of the jets gives key
information on the second derivative of the underlying shape of
the valence pion wavefunction
\cite{Frankfurt:1993it,Frankfurt:2000jm,BHDP}.  The diffractive
nuclear amplitude extrapolated to $t = 0$ should be linear in
nuclear number $A$ if color transparency is correct.  The
integrated diffractive rate should then scale as $A^2/R^2_A \sim
A^{4/3}$.

The results of a diffractive dijet dissociation experiment of this
type E791 at Fermilab using 500 GeV incident pions on nuclear
targets \cite{Aitala:2001hc} appear to be consistent with color
transparency.  The measured longitudinal momentum distribution of
the jets \cite{Aitala:2001hb} is consistent with a pion
light-front wavefunction of the pion with the shape of the
asymptotic distribution amplitude, $\phi^{\rm asympt}_\pi (x) =
\sqrt 3 f_\pi x(1-x)$.  Data from CLEO \cite{Gronberg:1998fj} for
the $\gamma \gamma^* \rightarrow \pi^0$ transition form factor
also favor a form for the pion distribution amplitude close to the
asymptotic solution to the perturbative QCD evolution equation
\cite{Lepage:1980fj}.

The interpretation of the diffractive dijet processes as measures
of the hadron distribution amplitudes has recently been questioned
by Braun {\em et al.} \cite{Braun:2001ih} and by Chernyak
\cite{Chernyak:2001ph} who have calculated the hard scattering
amplitude for such processes at next-to-leading order.  However,
these analyses neglect the integration over the transverse
momentum of the valence quarks and thus miss the logarithmic
ordering which is required for factorization of the distribution
amplitude and color-filtering in nuclear targets.

As noted above, the diffractive dissociation of a hadron or
nucleus can also occur via the Coulomb dissociation of a beam
particle on an electron beam (\eg\ at HERA or eRHIC) or on the
strong Coulomb field of a heavy nucleus (\eg\ at RHIC or nuclear
collisions at the LHC) \cite{BHDP}.  The amplitude for Coulomb
exchange at small momentum transfer is proportional to the first
derivative $\sum_i e_i {\partial \over \vec k_{T i}} \psi$ of the
light-front wavefunction, summed over the charged constituents.
The Coulomb exchange reactions fall off less fast at high
transverse momentum compared to pomeron exchange reactions since
the light-front wavefunction is effective differentiated twice in
two-gluon exchange reactions.

It will also be interesting to study diffractive tri-jet
production using proton beams $ p A \rightarrow {\rm Jet}_1 + {\rm
Jet}_2 + {\rm Jet}_3 + A^\prime $ to determine the fundamental
shape of the 3-quark structure of the valence light-front
wavefunction of the nucleon at small transverse separation
\cite{Frankfurt:1993it}. For example, consider the Coulomb
dissociation of a high energy proton at HERA.  The proton can
dissociate into three jets corresponding to the three-quark
structure of the valence light-front wavefunction.  We can demand
that the produced hadrons all fall outside an opening angle
$\theta$ in the proton's fragmentation region. Effectively all of
the light-front momentum $\sum_j x_j \simeq 1$ of the proton's
fragments will thus be produced outside an ``exclusion cone".
This then limits the invariant mass of the contributing Fock state
${\cal M}^2_n > \Lambda^2 = P^{+2} \sin^2\theta/4$ from below, so
that perturbative QCD counting rules can predict the fall-off in
the jet system invariant mass $\cal M$.  At large invariant mass
one expects the three-quark valence Fock state of the proton to
dominate.  The segmentation of the forward detector in azimuthal
angle $\phi$ can be used to identify structure and correlations
associated with the three-quark light-front wavefunction
\cite{BHDP}. An interesting possibility is that the distribution
amplitude of the $\Delta(1232)$ for $J_z = 1/2, 3/2$ is close to
the asymptotic form $x_1 x_2 x_3$,  but that the proton
distribution amplitude is more complex. This ansatz can also be
motivated by assuming a quark-diquark structure of the baryon
wavefunctions.  The differences in shapes of the distribution
amplitudes could explain why the $p \to\Delta$ transition form
factor appears to fall faster at large $Q^2$ than the elastic $p
\to p$ and the other $p \to N^*$ transition form factors
\cite{Stoler:1999nj}. One can use also measure the dijet structure
of real and virtual photons beams $ \gamma^* A \rightarrow {\rm
Jet}_1 + {\rm Jet}_2 + A^\prime $ to measure the shape of the
light-front wavefunction for transversely-polarized and
longitudinally-polarized virtual photons.  Such experiments will
open up a direct window on the amplitude structure of hadrons at
short distances. The light-front formalism is also applicable to
the description of nuclei in terms of their nucleonic and mesonic
degrees of freedom \cite{Miller:1999mi,Miller:2000ta}.
Self-resolving diffractive jet reactions in high energy
electron-nucleus collisions and hadron-nucleus collisions at
moderate momentum transfers can thus be used to resolve the
light-front wavefunctions of nuclei.

The first tests of color transparency involved large momentum
transfer quasi-elastic scattering processes in nuclei. Such
reactions are predicted in perturbative QCD to depend on the
scattering of small impact size hadron wavefunction configurations
\cite{Brodsky:1988xz}.  The onset of  color transparency in
proton-proton scattering in nuclei was first seen by Experiment
E834 at BNL by observing a rise in the ratio of quasi-elastic to
elastic $pp$ scattering at large angles and energies up to $\sqrt
s \sim 5$ GeV~\cite{Carroll:rp}.  Quasi-elastic proton-proton
scattering is advantageous over the analogous electron-proton
scattering reaction since the wavefunctions of the incoming and
outgoing hadron in high energy proton reactions would not suffer
rapid expansion.  However, E834 also revealed another remarkable
feature of quasi-elastic $pp$ scattering: the quenching of color
transparency at the largest measured energy measured by E834, in
direct contradiction to the predictions perturbative QCD.  A more
recent experiment, E850, using the EVA spectrometer has now
confirmed this unexpected effect through new measurements of the
transparency ratio at higher energies~\cite{Leksanov:2001ui}.

The quenching of color transparency observed in the E834 and E850
experiments is almost as important discovery as color transparency
itself.  It signals a nonperturbative effect in QCD which clearly
must be understood.  The quenching occurs at the center-of-mass
energy of $5$ GeV where the $pp$ elastic cross section also
displays another remarkable effect: the rate of scattering where
the spins of the initial protons are parallel and normal to the
scattering plane grows rapidly and becomes approximately $4$ times
as large as the spin-antiparallel rate~\cite{Court:1986dh}.   De
Teramond and I~\cite{Brodsky:1987xw} have noted that both
phenomenon occur just at the threshold for open charm hadron
production.  We have shown in fact that resonance production in
$pp$ elastic scattering due to a $uud uud c \bar c$ spin-1
resonance will in fact lead to a remarkably large spin correlation
$A_{NN}$ and quenching of color transparency above the charm
threshold.  If this explanation is validated (by the observation
of a significant open charm cross section near $5$ GeV center of
mass energy), then  E834 and E850 will have provided the first
evidence for an exotic QCD state with hidden charm.

\section{Higher Fock States and the Intrinsic Sea}

Since a hadronic wavefunction describes states off of the
light-front energy shell, there is a finite probability of the
projectile having fluctuations containing extra quark-antiquark
pairs, such as intrinsic strangeness charm, and bottom. In
contrast to the quark pairs arising from gluon splitting,
intrinsic quarks are multiply-connected to the valence quarks and
are thus part of the dynamics of the hadron.

Recently Franz, Polyakov, and Goeke have analyzed the properties
of the intrinsic heavy-quark fluctuations in hadrons using the
operator-product expansion~\cite{Franz:2000ee}. For example, the
light-cone momentum fraction carried by intrinsic heavy quarks in
the proton $x_{Q \bar Q}$ as measured by the $T^{+ + }$ component
of the energy-momentum tensor is related in the heavy-quark limit
to the forward matrix element $\langle p \vert {\hbox{tr}_c}
{(G^{+\alpha} G^{+ \beta} G_{\alpha \beta})/ m_Q^2 }\vert p
\rangle ,$ where $G^{\mu \nu}$ is the gauge field strength tensor.
Diagrammatically, this can be described as a heavy quark loop in
the proton self-energy with four gluons attached to the light,
valence quarks. Since the non-Abelian commutator $[A_\alpha,
A_\beta]$ is involved, the heavy quark pairs in the proton
wavefunction are necessarily in a color-octet state. It follows
from dimensional analysis that the momentum fraction carried by
the $Q\bar Q$ pair scales as $k^2_\perp / m^2_Q$ where $k_\perp$
is the typical momentum in the hadron wave function.  [In
contrast, in the case of Abelian theories, the contribution of an
intrinsic, heavy lepton pair to the bound state's structure first
appears in ${ O}(1/m_L^4)$.  One relevant operator corresponds to
the Born-Infeld $(F_{\mu\nu})^4$ light-by-light scattering
insertion, and the momentum fraction of heavy leptons in an atom
scales as $k^4_\perp / m_L^4$.]

The intrinsic sea is thus sensitive to the hadronic bound-state
structure \cite{Brodsky:1981se,Brodsky:1980pb}. The maximal
contribution of an intrinsic heavy quark occurs at $x_Q \simeq
{m_{\perp Q}/ \sum_i m_\perp}$ where $m_\perp =
\sqrt{m^2+k^2_\perp}$; \ie\ at large $x_Q$, since this minimizes
the invariant mass $\M^2_n$. The measurements of the charm
structure function by the EMC experiment are consistent with
intrinsic charm at large $x$ in the nucleon with a probability of
order $0.6 \pm 0.3 \% $ \cite{Harris:1996jx} which is consistent
with the recent estimates based on instanton fluctuations
\cite{Franz:2000ee}.

Thus one can identify two contributions to the heavy quark sea,
the ``extrinsic'' contributions which correspond to ordinary gluon
splitting, and the ``intrinsic" sea which is multi-connected via
gluons to the valence quarks. Intrinsic charm can be materialized
by diffractive dissociation into open or hidden charm states such
as $p p \to J/\psi X p', \Lambda_c X p'$. At HERA one can measure
intrinsic charm in the proton by Coulomb dissociation: $p e \to
\Lambda_C X e',$ and $J/\psi X e'.$  Since the intrinsic heavy
quarks tend to have the same rapidity as that of the projectile,
they are produced at large $x_F$ in the beam fragmentation region.

The presence of intrinsic charm quarks in the $B$ wave function
provides new mechanisms for $B$ decays. The characteristic momenta
characterizing the $B$ meson is most likely higher by a factor of
2 compared to the momentum scale of light mesons, This effect is
analogous to the higher momentum scale of muonium $\mu^+ e^-$
versus that of positronium $e^+ e^-$ in atomic physics because of
the larger reduced mass.  Thus one can expect a higher probability
for intrinsic charm in heavy hadrons compared to light hadrons.
For example, Chang and Hou have considered the production of final
states with three charmed quarks such as $B \to J/\psi D \pi$ and
$B \to J/\psi D^*$~\cite{Chang:2001iy}; these final states are
difficult to realize in the valence model, yet they occur
naturally when the $b$ quark of the intrinsic charm Fock state
$\ket{ b \bar u c \bar c}$ decays via $b \to c \bar u d$.  In
fact, the $J/\psi$ spectrum for inclusive $B \to J/\psi X$ decays
measured by CLEO and Belle shows a distinct enhancement at the low
$J/\psi$ momentum where such decays would kinematically occur.
Alternatively, this excess could reflect the opening of baryonic
channels such as $B \to J/\psi \bar p
\Lambda$~\cite{Brodsky:1997yr}.

Recently, Susan Gardner and I have shown that the presence of
intrinsic charm in the hadrons' light-front wave functions, even
at a few percent level, provides new, competitive decay mechanisms
for $B$ decays which are nominally
CKM-suppressed~\cite{Brodsky:2001yt}.  For example, the weak
decays of the $B$-meson to two-body exclusive states consisting of
strange plus light hadrons, such as $B \to \pi K$, are expected to
be dominated by penguin contributions since the tree-level $b\to s
u{\overline u}$ decay is CKM suppressed. However, higher Fock
states in the $B$ wave function containing charm quark pairs can
mediate the decay via a CKM-favored $b\to s c{\overline c}$
tree-level transition. Such intrinsic charm contributions can be
phenomenologically significant.  Since they mimic the amplitude
structure of ``charming'' penguin contributions
\cite{Colangelo:1989gi,Isola:2001bn,Isola:2001ar,Ciuchini:2001gv},
charming penguins need not be penguins at
all~\cite{Brodsky:2001yt}.

One can also distinguish ``intrinsic gluons" \cite{Brodsky:1990db}
which are associated with multi-quark interactions and extrinsic
gluon contributions associated with quark substructure.  One can
also use this framework to isolate the physics of the anomaly
contribution to the Ellis-Jaffe sum rule \cite{Bass:1999rn}. Thus
neither gluons nor sea quarks are solely generated by DGLAP
evolution, and one cannot define a resolution scale $Q_0$ where
the sea or gluon degrees of freedom can be neglected.

It is usually assumed that a heavy quarkonium state such as the
$J/\psi$ always decays to light hadrons via the annihilation of
its heavy quark constituents to gluons.  However, as Karliner and
I \cite{Brodsky:1997fj} have shown, the transition $J/\psi \to
\rho \pi$ can also occur by the rearrangement of the $c \bar c$
from the $J/\psi$ into the $\ket{ q \bar q c \bar c}$ intrinsic
charm Fock state of the $\rho$ or $\pi$.  On the other hand, the
overlap rearrangement integral in the decay $\psi^\prime \to \rho
\pi$ will be suppressed since the intrinsic charm Fock state
radial wavefunction of the light hadrons will evidently not have
nodes in its radial wavefunction.  This observation provides a
natural explanation of the long-standing
puzzle~\cite{Brodsky:1987bb} why the $J/\psi$ decays prominently
to two-body pseudoscalar-vector final states, breaking hadron
helicity conservation~\cite{Brodsky:1981kj}, whereas the
$\psi^\prime$ does not.

The higher Fock state of the proton $\ket{u u d s \bar s}$ should
resemble a $\ket{ K \Lambda}$ intermediate state, since this
minimizes its invariant mass $\M$.  In such a state, the strange
quark has a higher mean momentum fraction $x$ than the $\bar s$
\cite{Burkardt:1992di,Signal:1987gz,Brodsky:1996hc}.  Similarly,
the helicity of the intrinsic strange quark in this configuration
will be anti-aligned with the helicity of the nucleon
\cite{Burkardt:1992di,Brodsky:1996hc}.  This $Q \leftrightarrow
\bar Q$ asymmetry is a striking feature of the intrinsic
heavy-quark sea.

\section{Non-Perturbative Solutions of Light-Front Quantized QCD}

Is there any hope of computing light-front wavefunctions from
first principles?  The solution of the light-front Hamiltonian
equation $ H^{QCD}_{LC} \ket{\Psi} = M^2 \ket{\Psi}$ is an
eigenvalue problem which in principle determines the masses
squared of the entire bound and continuum spectrum of QCD.  If one
introduces periodic or anti-periodic boundary conditions, the
eigenvalue problem is reduced to the diagonalization of a discrete
Hermitian matrix representation of $H^{QCD}_{LC}.$ The light-front
momenta satisfy $x^+ = {2 \pi \over L} n_i$ and $P^+ = {2\pi \over
L} K$, where $\sum_i n_i = K.$ The number of quanta in the
contributing Fock states is restricted by the choice of harmonic
resolution.  A cutoff on the invariant mass of the Fock states
truncates the size of the matrix representation in the transverse
momenta. This is the essence of the DLCQ method
\cite{Pauli:1985ps}, which has now become a standard tool for
solving both the spectrum and light-front wavefunctions of
one-space one-time theories---virtually any $1+1$ quantum field
theory, including ``reduced QCD" (which has both quark and gluonic
degrees of freedom) can be completely solved using DLCQ
\cite{Dalley:1993yy,Antonuccio:1995fs}. The method yields not only
the bound-state and continuum spectrum, but also the light-front
wavefunction for each eigensolution
\cite{Antonuccio:1996hv,Antonuccio:1996rb}.

Dalley \etal\ have shown how one can use DLCQ in one space-one
time, with a transverse lattice to solve mesonic and gluonic
states in $ 3+1$ QCD \cite{Dalley:2000ii}. The spectrum obtained
for gluonium states is in remarkable agreement with lattice gauge
theory results, but with a huge reduction of numerical effort.
Hiller and I \cite{Hiller:1999cv} have shown how one can use DLCQ
to compute the electron magnetic moment in QED without resort to
perturbation theory.

There has been recent progress developing the computational tools
and renormalization methods which can make DLCQ a viable
computational method for QCD in physical space-time. John Hiller,
Gary McCartor, and I~\cite{Brodsky:2001ja} have shown how DLCQ can
be used to solve (3+1) theories and  obtain the spectrum and
light-front wavefunctions  of the bound state solutions despite
the large numbers of degrees of freedom needed to enumerate the
Fock basis.  A key feature of our work is the introduction of
Pauli Villars fields \cite{PauliVillars} in the DLCQ basis which
regulate the UV divergences and perform renormalization while
preserving the frame-independence of the theory
\cite{Brodsky:1998hs,Brodsky:1999xj}.

A recent application of DLCQ and Pauli Villars regularization to a
(3+1) quantum field theory with Yukawa interactions is given in
Ref.~\cite{Brodsky:2001ja}.  Only one heavy fermion is allowed in
the Fock states.  We include an additional effective interaction
which represents the contribution of the missing Z graph and
cancels an infrared singularity introduced by the instantaneous
fermion interaction. Cancellation of ultraviolet infinities is
then arranged by choosing imaginary couplings or an indefinite
metric. In our most recent work we used three heavy scalars, two
of which have negative norm.

In DLCQ, all light-front momentum variables are discretized, with
$p^+\rightarrow n\pi/L$ and $\vec{p}_\perp\rightarrow
\vec{n}_\perp\pi/L_\perp$, in terms of longitudinal and transverse
length scales $L$ and $L_\perp$. The total longitudinal momentum
is $P^+=K\pi/L$ and momentum fractions are given by $x=n/K$.  Wave
functions and the mass eigenvalue problem, where $H_{\rm
LC}=P^+P^-$, are naturally expressed in terms of momentum
fractions and the resolution $K$.  Hence $L$ disappears, and $K$
effectively takes its place as the resolution scale. The
transverse scale $L_\perp$ is set by a momentum cutoff and a
transverse resolution.  The integrals over wave functions which
make up the mass eigenvalue problem $H_{\rm LC}\Phi=M^2\Phi$ are
then approximated by the trapezoidal quadrature rule.  This yields
a matrix eigenvalue problem which is typically quite large but
also quite sparse.  Lanczos techniques~\cite{Lanczos} are used to
extract eigenvalues and eigenvectors for the lowest states, even
in the case of an indefinite metric~\cite{Brodsky:2001ja}.

The mass $M$ of the dressed single-fermion state is held fixed.
This is imposed by rearranging the mass eigenvalue problem into an
eigenvalue problem for the quantity $\delta M^2$:
\begin{eqnarray}
\lefteqn{x\left[M^2- \frac{M^2+p_\perp^2}{x}
   -\sum_j\frac{\mu_j^2+q_{\perp j}^2}{y_j}\right] \tilde{\phi}} \\
&&-\int\prod_j dy'_j d^2q'_{\perp j}\sqrt{xx'}{\cal
K}\tilde{\phi}' =\delta M^2\tilde{\phi}\,,  \nonumber
\end{eqnarray}
where ${\cal K}$ represents the original kernel and amplitudes are
related by $\phi=\sqrt{x}\tilde{\phi}$.

The coupling $g$ is constrained by imposing a condition on the
boson occupation number: $\langle :\!\!\phi^2(0)\!\!:\rangle
\equiv\Phi_\sigma^\dagger\!:\!\!\phi^2(0)\!\!:\!\Phi_\sigma$. This
quantity can be computed fairly efficiently as the sum
\begin{eqnarray}
\lefteqn{\langle :\!\!\phi^2(0)\!\!:\rangle
        =\sum_{n_i=0}^\infty \int \prod_j^{n_{\rm tot}}
                          \,dq_j^+d^2q_{\perp j} \sum_s (-1)^{(n_i)}}
    \\
  & & \times \left(\sum_{k=1}^n \frac{2}{q_k^+/P^+}\right)
       \left|\phi_{\sigma s}^{(n_i)}(\underline{q}_j;
       \underline{P}-\sum_j\underline{q}_j)\right|^2\,.
\nonumber
\end{eqnarray}
The constraint on $\langle :\!\!\phi^2(0)\!\!:\rangle$ can be
satisfied by solving it simultaneously with the eigenvalue
problem.

With the parameters fixed, we can compute various quantities, such
as the parton wavefunctions and momentum distributions, the form
factor slope at zero momentum transfer, the average numbers of
constituents, and the average constituent momenta. A
representative plot of the bosonic structure function
\begin{eqnarray}
\lefteqn{f_B(y)\equiv\sum_{n_i=0}^\infty\sum_s
   \int\,\prod_jdq_j^+d^2q_{\perp j} (-1)^{(n_i)}\sum_{k=1}^{n_0}}
\nonumber  \\
   &&\times  \delta(y-q_k^+/P^+)
      \left|\phi_{\sigma s}^{(n_i)}(\underline{q}_j;
                      \underline{P}-\sum_i\underline{q}_j)\right|^2\,,
\end{eqnarray}
is given in Fig.~\ref{fig:StructFn}.
\begin{figure}
\begin{center}
\includegraphics[width=9.5cm]{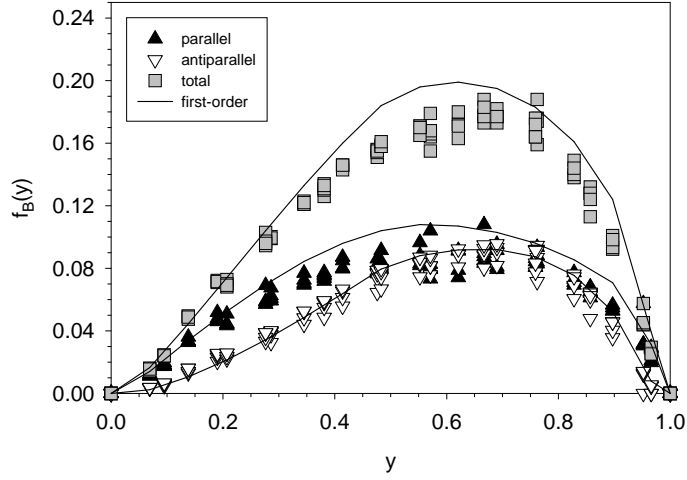}
\end{center}
\caption{\label{fig:StructFn} The boson structure function $f_B$
at various numerical resolutions for
$\langle:\!\!\phi^2(0)\!\!:\rangle=0.5$, with $M=\mu$, cutoff
$\Lambda^2=50\mu^2$, and Pauli--Villars masses $\mu_1^2=10\mu^2$,
$\mu_2^2=20\mu^2$, and $\mu_3^2=30\mu^2$.  The solid line is from
first-order perturbation theory. }
\end{figure}
We have also worked at somewhat stronger couplings where
deviations from first order perturbation theory becomes apparent;
however, high resolution is required, with $K=21$ to 39 and as
many as 15 transverse momentum points.  This resolution could be
achieved by limiting the number of constituents to 3, after
verifying that the contribution from higher sectors was
sufficiently small.

The success of application of DLCQ to the Yukawa theory with
Pauli-Villars regularization is encouraging.  One can compute
masses and wave functions for eigenstates for quantum field
theories in physical space-time. Another procedure, now under
investigation, is to use one heavy scalar and one heavy fermion,
both with negative norm, as suggested by the work of Paston {\em
et al}.~\cite{Paston} This method has the advantage of being free
of instantaneous fermion interactions.  An alternative and
interesting regularization is to apply DLCQ to finite
supersymmetric 3+1 theories, and then introduce supersymmetric
breaking.

We plan to continue to be explored these possibilities with
various model theories, leading eventually to the direct
application to QCD(3+1).  In fact, Paston {\em et
al}.~\cite{PastonQCD} have already obtained a PV-like
regularization of QCD which could, in principle, be solved by
DLCQ; however, given present computing power the number of fields
is possibly too large for meaningful calculations.

One can also formulate DLCQ so that supersymmetry is exactly
preserved in the discrete approximation, thus combining the power
of DLCQ with the beauty of supersymmetry
\cite{Antonuccio:1999ia,Lunin:1999ib,Haney:2000tk}. The ``SDLCQ"
method has been applied to several interesting supersymmetric
theories, to the analysis of zero modes, vacuum degeneracy,
massless states, mass gaps, and theories in higher dimensions, and
even tests of the Maldacena conjecture \cite{Antonuccio:1999ia}.
Broken supersymmetry is interesting in DLCQ, since it may serve as
a method for regulating non-Abelian theories
\cite{Brodsky:1999xj}.

There are also many possibilities for obtaining approximate
solutions of light-front wavefunctions in QCD.  QCD sum rules,
lattice gauge theory moments, and QCD inspired models such as the
bag model, chiral theories, provide important constraints. Guides
to the exact behavior of LC wavefunctions in QCD can also be
obtained from analytic or DLCQ solutions to toy models such as
``reduced" $QCD(1+1).$ The light-front and many-body Schr\"odinger
theory formalisms must match In the nonrelativistic limit.

It would be interesting to see if light-front wavefunctions can
incorporate chiral constraints such as soliton (Skyrmion) behavior
for baryons and other consequences of the chiral limit in the soft
momentum regime.  Solvable theories such as $QCD(1+1)$ are also
useful for understanding such phenomena.  It has been shown that
the anomaly contribution for the $\pi^0\to \gamma \gamma$ decay
amplitude is satisfied by the light-front Fock formalism in the
limit where the mass of the pion is light compared to its size
\cite{Lepage:1982gd}.

\section{Non-Perturbative Calculations of the Pion Distribution Amplitude}

The distribution amplitude $\phi(x,\widetilde Q)$ can be computed
from the integral over transverse momenta of the renormalized
hadron valence wavefunction in the light-cone gauge at fixed
light-front time \cite{Brodsky:1989pv}:
\begin{equation}
\phi(x,\widetilde Q) = \int d^2\vec{k_\perp}\thinspace \theta
\left({\widetilde Q}^2 - {\vec{k_\perp}^2\over x(1-x)}\right)
\psi^{(\widetilde Q)}(x,\vec{k_\perp}), \label{quarkdistamp}
\end{equation}
where a global cutoff in invariant mass is identified with the
resolution $\tilde Q$.  The distribution amplitude $\phi(x, \tilde
Q)$ is boost and gauge invariant and evolves in $\ln \tilde Q$
through an evolution equation
\cite{Lepage:1979za,Lepage:1979zb,Lepage:1980fj}. Since it is
formed from the same product of operators as the non-singlet
structure function, the anomalous dimensions controlling
$\phi(x,Q)$ dependence in the ultraviolet $\log Q$ scale are the
same as those which appear in the DGLAP evolution of structure
functions \cite{Brodsky:1980ny}. The decay $\pi \to \mu \nu$
normalizes the wave function at the origin: ${a_0/ 6} = \int^1_0
dx \phi(x,Q) = {f_\pi/ (2 \sqrt 3)}.$ One can also compute the
distribution amplitude from the gauge invariant Bethe-Salpeter
wavefunction at equal light-front time.  This also allows contact
with both QCD sum rules and lattice gauge theory; for example,
moments of the pion distribution amplitudes have been computed in
lattice gauge theory
\cite{Martinelli:1987si,Daniel:1991ah,DelDebbio:2000mq}.

Dalley \cite{Dalley:2000dh} has recently calculated the pion
distribution amplitude from QCD using a combination of the
discretized DLCQ method for the $x^-$ and $x^+$ light-front
coordinates with the transverse lattice method
\cite{Bardeen:1976tm,Burkardt:1996gp} in the transverse
directions,  A finite lattice spacing $a$ can be used by choosing
the parameters of the effective theory in a region of
renormalization group stability to respect the required gauge,
Poincar\'e, chiral, and continuum symmetries. The overall
normalization gives $f_{\pi} = 101$ MeV compared with the
experimental value of $93$ MeV. Figure \ref{Fig:DalleyCleo} (a)
compares the resulting DLCQ/transverse lattice pion wavefunction
with the best fit to the diffractive di-jet data (see the next
section) after corrections for hadronization and experimental
acceptance \cite{Ashery:1999nq}. The theoretical curve is somewhat
broader than the experimental result.  However, there are
experimental uncertainties from hadronization and theoretical
errors introduced from finite DLCQ resolution, using a nearly
massless pion, ambiguities in setting the factorization scale
$Q^2$, as well as errors in the evolution of the distribution
amplitude from 1 to $10~{\rm GeV}^2$.  Instanton models also
predict a pion distribution amplitude close to the asymptotic form
\cite{Petrov:1999kg}. In contrast,  recent lattice results from
Del Debbio {\em et al.} \cite{DelDebbio:2000mq} predict a much
narrower shape for the pion distribution amplitude than the
distribution predicted by the transverse lattice. A new result for
the proton distribution amplitude treating nucleons as chiral
solitons has recently been derived by Diakonov and Petrov
\cite{Diakonov:2000pa}. Dyson-Schwinger models \cite{Hecht:2000xa}
of hadronic Bethe-Salpeter wavefunctions can also be used to
predict light-front wavefunctions and hadron distribution
amplitudes by integrating over the relative $k^-$ momentum.  There
is also the possibility of deriving Bethe-Salpeter wavefunctions
within light-front gauge quantized QCD \cite{Srivastava:2000gi} in
order to properly match to the light-cone gauge Fock state
decomposition.

\section{Calculating and Modelling Light-Front Wavefunctions}

Many features of the light-front wavefunctions follow from general
arguments. Light-front wavefunctions satisfy the equation of
motion:
$$ H^{QCD}_{LC} \ket{\Psi} = (H^{0}_{LC} + V_{LC} )\ket{\Psi} = M^2
\ket{\Psi},$$ which has the Heisenberg matrix form in Fock space:
$$\large [M^2 - \sum_{i=1}^n{m_{\perp i}^2\over x_i} \large ]\psi_n =
\sum_{n'}\int \VEV{n|V|n'} \psi_{n'}$$
where the convolution and sum is understood over the Fock number,
transverse momenta, plus momenta and helicity of the intermediate
states.  Here $m^2_\perp = m^2 + k^2_\perp.$ Thus, in general,
every light-front Fock wavefunction has the form:
$$\psi_n={\Gamma_n\over M^2-\sum_{i=1}^n{m_{\perp i}^2\over x_i}}$$
where $\Gamma_n = \sum_{n'}\int V_{n {n'}} \psi_n$. The main
dynamical dependence of a light-front wavefunction away from the
extrema is controlled by its light-front energy denominator.  The
maximum of the wavefunction occurs when the invariant mass of the
partons is minimal; \ie, when all particles have equal rapidity
and are all at rest in the rest frame. In fact, Dae  Sung Hwang
and I \cite{BH} have noted that one can rewrite the wavefunction
in the form:
$$\psi_n= {\Gamma_n\over M^2
[\sum_{i=1}^n {(x_i-{\hat x}_i)^2\over x_i} + \delta^2]}$$
where $x_i = {\hat x}_i\equiv{m_{\perp i}/ \sum_{i=1}^n m_{\perp
i}}$ is the condition for minimal rapidity differences of the
constituents.  The key parameter is $ M^2-\sum_{i=1}^n{m_{\perp
i}^2/ {\hat x}_i}\equiv -M^2\delta^2.$ We can also interpret
$\delta^2 \simeq 2 \epsilon / M $ where $ \epsilon = \sum_{i=1}^n
m_{\perp i}-M $ is the effective binding energy. This form shows
that the wavefunction is a quadratic form around its maximum, and
that the width of the distribution in $(x_i - \hat x_i)^2$ (where
the wavefunction falls to half of its maximum) is controlled by
$x_i \delta^2$ and the transverse momenta $k_{\perp_i}$.  Note
also that the heaviest particles tend to have the largest $\hat
x_i,$ and thus the largest momentum fraction of the particles in
the Fock state, a feature familiar from the intrinsic charm model.
For example, the $b$ quark has the largest momentum fraction at
small $k_\perp$ in the $B$ meson's valence light-front
wavefunction,, but the distribution spreads out to an
asymptotically symmetric distribution around $x_b \sim 1/2$ when
$k_\perp \gg m^2_b.$

We can also discern some general properties of the numerator of
the light-front wavefunctions. $\Gamma_n(x_i, k_{\perp i},
\lambda_i)$. The transverse momentum dependence of $\Gamma_n$
guarantees $J_z$ conservation for each Fock state.  For example,
one of the three light-front Fock wavefunctions of a $J_z = +1/2$
lepton in QED perturbation theory is $
\psi^{\uparrow}_{+\frac{1}{2}\, +1} (x,{\vec
k}_{\perp})=-{\sqrt{2}} \frac{(-k^1+{\mathrm i} k^2)}{x(1-x)}\,
\varphi \ ,$ where $ \varphi=\varphi (x,{\vec k}_{\perp})=\frac{
e/\sqrt{1-x}}{M^2-({\vec k}_{\perp}^2+m^2)/x-({\vec
k}_{\perp}^2+\lambda^2)/(1-x)}\ . $ The orbital angular momentum
projection in this case is $\ell^z = -1.$ The spin structure
indicated by perturbative theory provides a template for the
numerator structure of the light-front wavefunctions even for
composite systems. The structure of the electron's Fock state in
perturbative QED shows that it is natural to have a negative
contribution from relative orbital angular momentum which balances
the $S_z$ of its photon constituents. We can also expect a
significant orbital contribution to the proton's $J_z$ since
gluons carry roughly half of the proton's momentum, thus providing
insight into the ``spin crisis" in QCD.

The high $x \to 1$ and high $k_\perp$ limits of the hadron
wavefunctions control processes and reactions in which the hadron
wavefunctions are highly stressed.  Such configurations involve
far-off-shell intermediate states and can be systematically
treated in perturbation theory
\cite{Brodsky:1995kg,Lepage:1980fj}. This leads to counting rule
behavior for the quark and gluon distributions at $x \to 1$.
Notice that $x\to 1$ corresponds to $k^z \to -\infty$ for any
constituent with nonzero mass or transverse momentum.

The above discussion suggests that an approximate form for the
hadron light-front wavefunctions could be constructed through
variational principles and by minimizing the expectation value of
$H^{QCD}_{LC}.$

\section{Structure Functions are Not Parton Distributions}

Ever since the earliest days of the parton model, it has been
assumed that the leading-twist structure functions $F_i(x,Q^2)$
measured in deep inelastic lepton scattering are determined by the
{\it probability} distribution of quarks and gluons as determined
by the light-front wavefunctions of the target.  For example, the
quark distribution is
$$
{ P}_{\qu/N}(x_B,Q^2)= \sum_n \int^{k_{i\perp}^2<Q^2}\left[
\prod_i\, dx_i\, d^2k_{\perp i}\right] |\psi_n(x_i,k_{\perp i})|^2
\sum_{j=q} \delta(x_B-x_j).
$$
The identification of structure functions with the square of
light-front wavefunctions is usually made in LC gauge $n\cdot A =
A^+=0$, where the path-ordered exponential in the operator product
for the forward virtual Compton amplitude apparently reduces to
unity. Thus the deep inelastic lepton scattering cross section
(DIS) appears to be fully determined by the probability
distribution of partons in the target. However, Paul Hoyer, Nils
Marchal, Stephane Peigne, Francesco Sannino, and I
\cite{Brodsky:2001ue} have recently shown that the leading-twist
contribution to DIS is affected by diffractive rescattering of a
quark in the target, a coherent effect which is not included in
the light-front wavefunctions, even in light-cone gauge.  The
distinction between structure functions and parton probabilities
is already implied by the Glauber-Gribov picture of nuclear
shadowing~\cite{Gribov:1969jf,Brodsky:1969iz,Brodsky:1990qz,Piller:2000wx}.
In this framework shadowing arises from interference between
complex rescattering amplitudes involving on-shell intermediate
states, as in Fig.~\ref{brodsky2}.  In contrast, the wave function
of a stable target is strictly real since it does not have on
energy-shell configurations.  A probabilistic interpretation of
the DIS cross section is thus precluded.

It is well-known that in Feynman and other covariant gauges one
has to evaluate the corrections to the ``handbag" diagram due to
the final state interactions of the struck quark (the line
carrying momentum $p_1$ in Fig. \ref{brodsky1}) with the gauge
field of the target.  In light-cone gauge, this effect also
involves rescattering of a spectator quark, the $p_2$ line in Fig.
\ref{brodsky1}.  The light-cone gauge is singular -- in
particular, the gluon propagator $ d_{LC}^{\mu\nu}(k) =
\frac{i}{k^2+\ieps}\left[-g^{\mu\nu}+\frac{n^\mu k^\nu+ k^\mu
n^\nu}{n\cdot k}\right] \label{lcprop} $ has a pole at $k^+ = 0$
which requires an analytic prescription.  In final-state
scattering involving on-shell intermediate states, the exchanged
momentum $k^+$ is of \order{1/\nu} in the target rest frame, which
enhances the second term in the propagator.  This enhancement
allows rescattering to contribute at leading twist even in LC
gauge.

\begin{figure}
\centering
\includegraphics[width=6in]{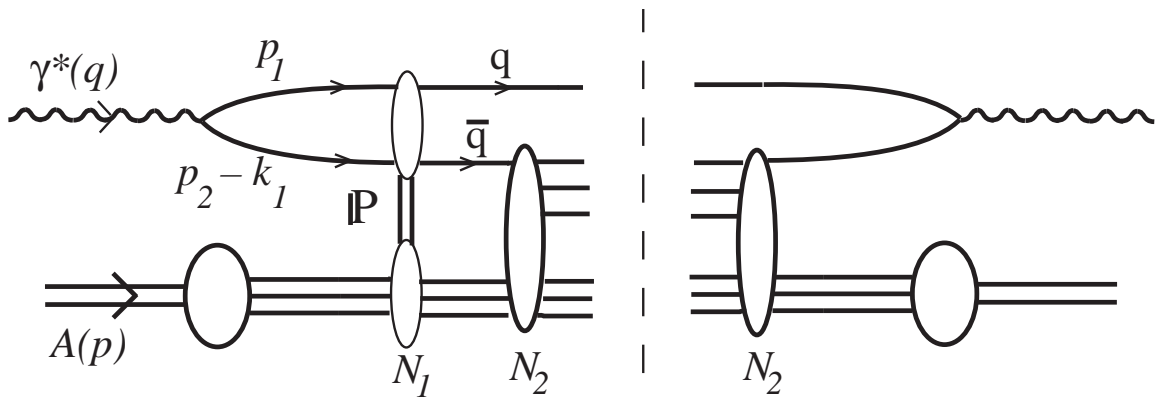}
\caption[*]{Glauber-Gribov shadowing involves interference between
rescattering amplitudes. \label{brodsky2}}
\end{figure}


The issues involving final state interactions even occur in the
simple framework of abelian gauge theory with scalar quarks.
Consider a frame with $q^+ < 0$. We can then distinguish FSI from
ISI using LC time-ordered perturbation
theory~\cite{Lepage:1980fj}. Figure~\ref{brodsky1} illustrates two
LCPTH diagrams which contribute to the forward $\gamma^* T \to
\gamma^* T$ amplitude, where the target $T$ is taken to be a
single quark. In the aligned jet kinematics the virtual photon
fluctuates into a \qu\qb\ pair with limited transverse momentum,
and the (struck) quark takes nearly all the longitudinal momentum
of the photon. The initial \qu\ and \qb\ momenta are denoted $p_1$
and $p_2-k_1$, respectively,

\begin{figure}
\includegraphics[width=6in]{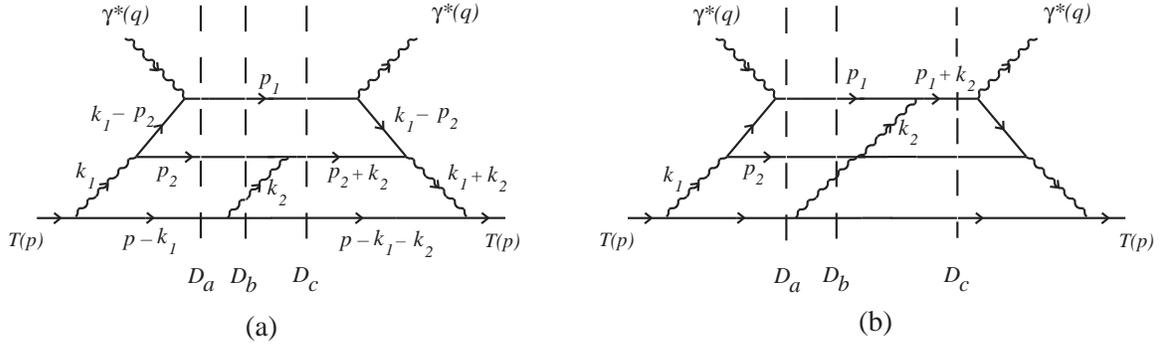}
\caption[*]{Two types of final state interactions.  (a) Scattering
of the antiquark ($p_2$ line), which in the aligned jet kinematics
is part of the target dynamics.  (b) Scattering of the current
quark ($p_1$ line).  For each LC time-ordered diagram, the
potentially on-shell intermediate states -- corresponding to the
zeroes of the denominators $D_a, D_b, D_c$ -- are denoted by
dashed lines.} \label{brodsky1}
\end{figure}

The calculation of the rescattering effect of DIS in Feynman and
light-cone gauge through three loops is given in detail in
Ref.~\cite{Brodsky:2001ue}.  The result can be resummed and is
most easily expressed in eikonal form in terms of transverse
distances $r_\perp, R_\perp$ conjugate to $p_{2\perp}, k_\perp$.
The deep inelastic cross section can be expressed as \beq
Q^4\frac{d\sigma}{dQ^2\, dx_B} =
\frac{\alpha}{16\pi^2}\frac{1-y}{y^2} \frac{1}{2M\nu} \int
\frac{dp_2^-}{p_2^-}\,d^2\rvec_\perp\, d^2\Rvec_\perp\, |\tilde
M|^2 \label{transcross} \eeq where \beq |\tilde{
M}(p_2^-,\rvec_\perp, \Rvec_\perp)| = \left|\frac{\sin \left[g^2\,
W(\rvec_\perp, \Rvec_\perp)/2\right]}{g^2\, W(\rvec_\perp,
\Rvec_\perp)/2} \tilde{A}(p_2^-,\rvec_\perp, \Rvec_\perp)\right|
\label{Interference} \eeq is the resummed result.  The Born
amplitude is \beq \tilde A(p_2^-,\rvec_\perp, \Rvec_\perp) = 2eg^2
M Q p_2^-\, V(m_\pl r_\perp) W(\rvec_\perp, \Rvec_\perp)
\label{Atildeexpr} \eeq where $ m_\pl^2 = p_2^-Mx_B + m^2
\label{mplus}$ and \beq V(m\, r_\perp) \equiv \int
\frac{d^2\pvec_\perp}{(2\pi)^2}
\frac{e^{i\rvec_\perp\cdot\pvec_{\perp}}}{p_\perp^2+m^2} =
\frac{1}{2\pi}K_0(m\,r_\perp) \label{Vexpr} \eeq The rescattering
effect of the dipole of the $q \bar q$ is controlled by \beq
W(\rvec_\perp, \Rvec_\perp) \equiv \int
\frac{d^2\kvec_\perp}{(2\pi)^2}
\frac{1-e^{i\rvec_\perp\cdot\kvec_{\perp}}}{k_\perp^2}
e^{i\Rvec_\perp\cdot\kvec_{\perp}} = \frac{1}{2\pi}
\log\left(\frac{|\Rvec_\perp+\rvec_\perp|}{R_\perp} \right).
\label{Wexpr} \eeq The fact that the coefficient of $\tilde A$ in
\eq{Interference} is less than unity for all $\rvec_\perp,
\Rvec_\perp$ shows that the rescattering corrections reduce the
cross section.  It is the analog of nuclear shadowing in our
model.

We have also found the same result for the deep inelastic cross
sections in light-cone gauge.  Three prescriptions for defining
the propagator pole at $k^+ =0$ have been used in the literature:
\beq \label{prescriptions} \frac{1}{k_i^+} \rightarrow
\left[\frac{1}{k_i^+} \right]_{\eta_i} = \left\{
\begin{array}{cc}
k_i^+\left[(k_i^+ -i\eta_i)(k_i^+ +i\eta_i)\right]^{-1} & ({\rm PV}) \\
\left[k_i^+ -i\eta_i\right]^{-1} & ({\rm K}) \\
\left[k_i^+ -i\eta_i \epsilon(k_i^-)\right]^{-1} & ({\rm ML})
\end{array} \right.
\eeq the principal-value, Kovchegov~\cite{Kovchegov:1997pc}, and
Mandelstam-Leibbrandt~\cite{Leibbrandt:1987qv} prescriptions. The
`sign function' is denoted $\epsilon(x)=\Theta(x)-\Theta(-x)$.
With the PV prescription we have $ I_{\eta} = \int dk_2^+
\left[\frac{1}{k_2^+} \right]_{\eta_2} = 0. $ Since an individual
diagram may contain pole terms $\sim 1/k_i^+$, its value can
depend on the prescription used for light-cone gauge. However, the
$k_i^+=0$ poles cancel when all diagrams are added;  the net is
thus prescription-independent, and it agrees with the Feynman
gauge result. It is interesting to note that the diagrams
involving rescattering of the struck quark $p_1$ do not contribute
to the leading-twist structure functions if we use the Kovchegov
prescription to define the light-cone gauge.  In other
prescriptions for light-cone gauge the rescattering of the struck
quark line $p_1$ leads to an infrared divergent phase factor $\exp
i\phi$: \beq \phi = g^2 \, \frac{I_{\eta}-1}{4 \pi} \, K_0(\lambda
R_{\perp}) + {{O}}(g^6) \eeq where $\lambda$ is an infrared
regulator, and $I_{\eta}= 1$ in the $K$ prescription. The phase is
exactly compensated by an equal and opposite phase from
final-state interactions of line $p_2$. This irrelevant change of
phase can be understood by the fact that the different
prescriptions are related by a residual gauge transformation
proportional to $\delta(k^+)$ which leaves the light-cone gauge
$A^+ = 0$ condition unaffected.

Diffractive contributions which leave the target intact thus
contribute at leading twist to deep inelastic scattering.  These
contributions do not resolve the quark structure of the target,
and thus they are contributions to structure functions which are
not parton probabilities. More generally, the rescattering
contributions shadow and modify the observed inelastic
contributions to DIS.

Our analysis in the $K$ prescription for light-cone gauge
resembles the ``covariant parton model" of Landshoff, Polkinghorne
and Short~\cite{Landshoff:1971ff,Brodsky:1973hm} when interpreted
in the target rest frame.  In this description of small $x$ DIS,
the virtual photon with positive $q^+$ first splits into the pair
$p_1$ and $p_2$.  The aligned quark $p_1$ has no final state
interactions.  However, the antiquark line $p_2$ can interact in
the target with an effective energy $\hat s \propto {k_\perp^2/x}$
while staying close to its mass shell.  Thus at small $x$ and
large $\hat s$, the antiquark $p_2$ line can first multiple
scatter in the target via pomeron and Reggeon exchange, and then
it can finally scatter inelastically or be annihilated. The DIS
cross section can thus be written as an integral of the
$\sigma_{\bar q p \to X}$ cross section over the $p_2$ virtuality.
In this way, the shadowing of the antiquark in the nucleus
$\sigma_{\bar q A \to X}$ cross section yields the nuclear
shadowing of DIS~\cite{Brodsky:1990qz}.  Our analysis, when
interpreted in frames with $q^+ > 0,$ also supports the color
dipole description of deep inelastic lepton scattering at small
$x$.  Even in the case of the aligned jet configurations, one can
understand DIS as due to the coherent color gauge interactions of
the incoming quark-pair state of the photon interacting first
coherently and finally incoherently in the target.

\section{A Light-Front Event Amplitude Generator}

The light-front formalism can be used as an ``event amplitude
generator" for high energy physics reactions where each particle's
final state is completely labelled in momentum, helicity, and
phase.  The application of the light-front time evolution operator
$P^-$ to an initial state systematically generates the tree and
virtual loop graphs of the $T$-matrix in light-front time-ordered
perturbation theory in light-cone gauge.  The loop integrals only
involve integrations over the momenta of physical quanta and
physical phase space $\prod d^2k_{\perp i} d k^+_i$.  Renormalized
amplitudes can be explicitly constructed by subtracting from the
divergent loops amplitudes with nearly identical integrands
corresponding to the contribution of the relevant mass and
coupling counter terms (the ``alternating denominator
method")~\cite{Brodsky:1973kb}.  The natural renormalization
scheme to use for defining the coupling in the event amplitude
generator is a physical effective charge such as the pinch
scheme~\cite{Cornwall:1989gv}. The argument of the coupling is
then unambiguous~\cite{Brodsky:1994eh}.  The DLCQ boundary
conditions can be used to discretize the phase space and limit the
number of contributing intermediate states without violating
Lorentz invariance. Since one avoids dimensional regularization
and nonphysical ghost degrees of freedom, this method of
generating events at the amplitude level could provide a simple
but powerful tool for simulating events both in QCD and the
Standard Model.

\section{The Light-Front Partition Function}

In the usual treatment of classical thermodynamics, one considers
an ensemble of particles $n = 1, 2, ... N$ which have energies
$\{E_n\}$ at a given ``instant" time $t$. The partition function
is defined as $Z = \sum_n \exp-{E_n\over kT}.$ Similarly, in
quantum mechanics, one defines a quantum-statistical partition
function as $Z = tr \exp{-\beta H}$ which sums over the
exponentiated-weighted energy eigenvalues of the system.

In the case of relativistic systems, it is natural to characterize
the system at a given light-front time $\tau = t +z/c$; i.e., one
determines the state of each particle in the ensemble as its
encounters the light-front. Thus we can define a light-front
partition function
$$Z_{LC} = \sum_n \exp -{p^-_n\over kT_{LC}}$$
by summing over the particles' light-front energies $p^- = p^0 -
p^z = {p^2_\perp + m^2 \over p^+}$.  The total momentum is $P^+ =
\sum p^+_n,$ $ \vec P_\perp = \sum_n \vec p_{\perp n}$, and the
total mass is defined from $P^+P^--P^2_\perp=M^2$.  The product
${M \over P^-} T_{LC}$ is boost invariant.  In the center of mass
frame where $\vec P =0$ and thus $P^+ = P^- = M$.  It is also
possible to consistently impose boundary conditions at fixed $x^-
= z - ct$ and $x_\perp$, as in DLCQ.  The momenta $p^+_n, \vec
p_{\perp n}$ then become discrete. The corresponding light-front
quantum-statistical partition function is $Z = tr \exp{-\beta_{LC}
H_{LC}}$ where $H_{LC} = i {\partial\over
\partial \tau}$ is the light-front Hamiltonian.

For non-relativistic systems the light-front partition function
reduces to the standard definition.  However, the light-front
partition function should be advantageous for analyzing
relativistic systems such as heavy ion collisions, since, like
true rapidity, $y = \ln {p^+\over P^+},$ light-front variables
have simple behavior under Lorentz boosts.  The light-front
formalism also takes into account the point that a phase
transition does not occur simultaneously in $t$, but propagates
through the system with a finite wave velocity.

\section*{Acknowledgments}
Work supported by the Department of Energy under contract number
DE-AC03-76SF00515. I wish to thank the organizers of this meeting,
A. Bialas, M. A. Nowak, and M. Sadzikowski  for their outstanding
hospitality in Zakopane.  Much of this work is based on
collaborations, particularly with Markus Diehl, Paul Hoyer, Dae
Sung Hwang,  Peter Lepage, Bo-Qiang Ma,  Hans Christian Pauli Ivan
Schmidt, and Prem Srivastava.

\end{document}